\documentclass[aps,prmaterials,amsmath,amssymb,reprint,superscriptaddress]{revtex4-1}
\usepackage{graphicx} 
\usepackage[version=4]{mhchem}

\begin{document}

\title{Deciphering structural and magnetic disorder in the chiral skyrmion host materials 
Co$_x$Zn$_y$Mn$_z$ ($x+y+z=20$)}

\author{Joshua D. Bocarsly}
\email{jdbocarsly@mrl.ucsb.edu}
\affiliation{Materials Department and Materials Research Laboratory, 
University of California, Santa Barbara, California 93106, United States}

\author{Colin Heikes}
\affiliation{Center for Neutron Research, National Institute of Standards and Technology,\\
Gaithersburg, Maryland 20899, United States} 

\author{Craig M. Brown}
\affiliation{Center for Neutron Research, National Institute of Standards and Technology,\\ 
Gaithersburg, Maryland 20899, United States} 

\author{Stephen D. Wilson}
\affiliation{Materials Department and Materials Research Laboratory, 
University of California, Santa Barbara, California 93106, United States} 

\author{Ram Seshadri}
\affiliation{Materials Department and Materials Research Laboratory, 
University of California, Santa Barbara, California 93106, United States}

\date{\today}
\begin{abstract}
Co$_x$Zn$_y$Mn$_z$ ($x + y + z =$ 20) compounds crystallizing in the chiral $\beta$-Mn crystal structure are known
to host skyrmion spin textures even at elevated temperatures. As in other chiral cubic skyrmion hosts, skyrmion lattices in these materials 
are found at equilibrium in a small pocket just below the magnetic Curie temperature. Remarkably, Co$_x$Zn$_y$Mn$_z$ compounds
have also been found to host metastable non-equilibrium skyrmion lattices in a broad temperature and field range,
including down to zero-field and low temperature. This behavior is believed to be related to disorder present in the materials.
Here, we use neutron and synchrotron diffraction, density functional theory calculations, and DC and AC magnetic measurements, to characterize the atomic and magnetic
disorder in these materials. We demonstrate that Co has a strong site-preference for the diamondoid 8c site in the crystal
structure, while Mn tends to share the geometrically frustrated 12d site with Zn, due to its ability to develop a large local moment on that site.
This magnetism-driven site specificity leads to distinct magnetic behavior for the Co-rich 8c sublattice and the Mn on the 12d sublattice. 
The Co-rich sublattice orders at high temperatures (compositionally tunable between 210\,K and 470\,K) with a moment around 1 $\mu_B$/atom and maintains this order to low temperature. The Mn-rich sublattice holds larger moments (about 3\,$\mu_B$) which remain fluctuating below the Co moment
ordering temperature. At lower temperature, the fluctuating Mn moments freeze into a reentrant disordered cluster-glass state with no net moment, while the Co moments maintain order. This two-sublattice behavior allows for the observed coexistence of strong magnetic disorder and ordered magnetic states such as helimagnetism and skyrmion lattices.

\end{abstract}
\maketitle

\section{Introduction}

Chiral magnetic nanostructures, including skyrmions and other long-wavelength modulated spin structures, have been 
the subject of intense and increasing research attention in the past decade. From an applications viewpoint, skyrmionic 
spin textures are exciting because they promise to enable magnetic racetrack memory and other kinds of low-power, high-density 
spintronics \cite{Fert2013}. From a scientific viewpoint, the discovery of skyrmion spin textures has revealed the tremendous diversity 
of magnetic nanostructures that can form spontaneously in seemingly simple magnetic
materials \cite{Muhlbauer2009,Yu2010,Yu2014,Kezsmarki2015,Phatak2016,Nayak2017}.

A goal in the search for new bulk skyrmion hosts has been to find materials that can exhibit these spin textures in a broad
temperature range around room temperature. In most bulk skyrmion hosts, stable skyrmion lattices are only observed in a 
narrow temperature and field pocket just below the magnetic Curie temperature \cite{Bauer2016, Munzer2010, Bauer2012, Bocarsly2018}. 
The stability of skyrmion lattices in this pocket is understood to arise from a combination of long-range magnetic interactions
and thermal fluctuations \cite{Muhlbauer2009,Buhrandt2013}. A true room-temperature skyrmion lattice in a bulk material was first 
reported in 2015 with the discovery of a hexagonal lattices of Bloch skyrmions in $\beta$-Mn structured Co$_x$Zn$_y$Mn$_z$ ($x+y+z=$20, $z \le 7$) \cite{Tokunaga2015} compounds. This family has Curie temperatures than can be compositionally tuned between about 210\,K and 470\,K, with each 
composition hosting skyrmions near its Curie temperature.

Beyond the equilibrium phase diagram, the topological protection of skyrmions implies the possibility of long-lived metastable skyrmions outside of 
the narrow pockets of stability. Indeed, by starting in the stable skyrmion lattice phase and quenching temperature, long-lived 
metastable skyrmion lattices have been observed in several bulk skyrmion hosts. In the chemically well-ordered material MnSi, 
cooling rates of hundreds of kelvins per second are needed to achieve this state \cite{Oike2015}. 
In a high-pressure cell, this state is achieved with moderate cooling rates \cite{Ritz2013}, a phenomenon that has been attributed
 to disorder induced by pressure inhomogeneities. Alternately, the intrinsically chemically disordered B20 compound
Fe$_{1-x}$Co$_{x}$Si exhibits long-lived metastable skyrmions with moderate cooling rates at ambient pressure \cite{Munzer2010}.

 \begin{figure}[t!]
	\centering
	\includegraphics[width=0.9\columnwidth]{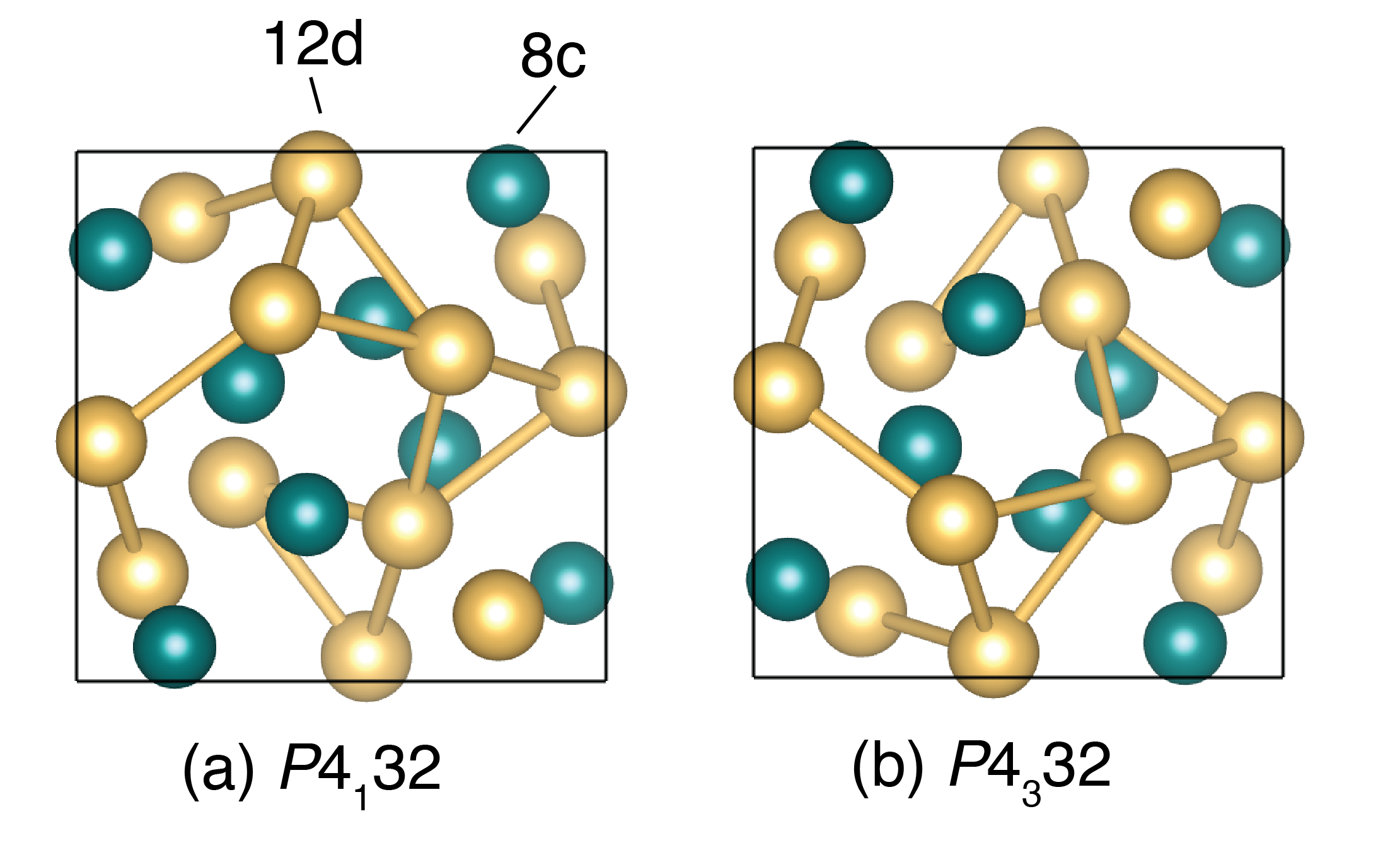}
	\caption{The $\beta$-Mn crystal structure, comprising 8c and 12d atomic sites, is displayed in its two
	 chiral enatiomers, $P4_132$ (a) and  $P4_332$ (b). The two structures differ only in handedness, and therefore should have the
	 same formation energy, with individual crystallites of a sample forming in one or the other configuration. The origin of each cell has been translated by 1/4 of a unit cell from the standard setting such that the chiral four-fold screw 
	axis ($4_1$ or $4_3$) is centered in the cube face. The connecting lines indicate 12d--12d contacts of 
	length $\approx$2.7\,\AA, showing how the 12d sublattice is constructed of equilateral triangles arranged in a 
	helix. The 8c sublattice, on the other hand, can be viewed as a distorted diamond lattice. }
	\label{fig:intro}
\end{figure}

As in Fe$_{1-x}$Co$_{x}$Si, a metastable skyrmion lattice may be formed in Co$_x$Zn$_y$Mn$_z$ by cooling through the
stable skyrmion lattice pocket at moderate rate. This metastable phase is observed in a broad temperature and field range, including down to zero field and from above room temperature down to at least 20\,K \cite{Karube2016,Karube2017}. At low temperatures and fields, the skyrmion lattice in \ce{Co8Zn8Mn4} goes through a reversible transition from hexagonal to square symmetry, while approximately maintaining the overall number of skyrmions.
Remarkably, this ordering appears to coexist with disordered spin glass behavior at low temperature \cite{Karube2018}. For a Mn-rich composition, \ce{Co7Zn7Mn6}, a second 
type of stable skyrmion lattice has recently been observed at low temperature in addition to the typical hexagonal lattice which is formed
near the Curie temperature \cite{Karube2018}. This low-temperature skyrmion lattice is disordered, and is suggested to be stabilized by short-range
frustrated magnetic fluctuations in the Co$_x$Zn$_y$Mn$_z$ magnetic ions, as opposed to the thermal fluctuations which stabilize the skyrmion
lattice near the Curie temperature. Given these observations, it is clear that the disorder in Co$_x$Zn$_y$Mn$_z$ drives new and interesting magnetic behavior. However, little has been done to characterize the atomic and magnetic disorder in this system, and understand its origins and couplings. 

A large amount of atomic and magnetic disorder is possible in the highly-flexible $\beta$-Mn structure, which is shown in Fig.\,\ref{fig:intro}. This structure consists of two crystallographic sites: a small eightfold site (8c) which forms a distorted diamondoid
network, and a larger twelvefold site (12d), which forms a geometrically frustrated hyperkagome lattice. $\beta$-Mn itself is proposed to show a spin liquid state \cite{Nakamura1997}, a property which is driven by geometric frustration in the 12d sublattice. Certain substitutions into 
$\beta$-Mn drive the system into a spin-glass with antiferromagnetic correlations \cite{Nakamura1997, Karlsen2009},
while nanoparticles of $\epsilon$-Co, which have $\beta$-Mn structure, appear to be 
ferromagnetic \cite{Graf2006}. Other compositions include the superconductors \ce{Li2(Pt/Pd)3B} \cite{Yuan2006} and \ce{Rh2Mo3N} \cite{Wei2016}. 

Co$_x$Zn$_y$Mn$_z$ compounds crystalize in this structure across a broad volume of the ternary phase diagram stretching
from $\beta$-Mn itself, to Mn$_{0.6}$Co$_{0.4}$ and CoZn \cite{Hori2007}. Due to the similar electronegativities and atomic 
radii of Mn, Co, and Zn, it is expected that the elements could mix on either of these two sites, leading to a material 
with a large amount of compositional disorder within the unit cell. Establishing the atomic distribution over the two sites is crucial to 
understanding the magnetic properties of this system, as ions on the two atomic sites are expected to have different magnetic properties
due to the different sizes and topologies of the two sites. In fact, in pure elemental $\beta$-Mn, Mn atoms on the 12d site are found to have much stronger local magnetic moments than on the 8c site \cite{Nakamura1997, Hama2004}.

In this contribution, we characterize the nature of the atomic and magnetic disorder in the skyrmion host materials Co$_x$Zn$_y$Mn$_z$ using neutron
and synchrotron diffraction, density functional theory calculations, and DC and AC magnetic measurements. The high-temperature diffraction and DFT results demonstrate
that the Co atoms have a site-preference for the 8c site, while Mn and Zn atoms tend to distribute randomly on the larger 12d site, 
which is geometrically frustrated. The Mn site preference is driven by the Mn atoms' ability to develop a large moment
(about 3 $\mu_B$) when sitting on this site. The magnetization measurements and low-temperature neutron diffraction reveal distinct magnetic
behavior between the Co sublattice and the Mn/Zn sublattice. The Co atoms order ferromagnetically at high temperature (between 100\,K and 470\,K, depending on composition) and remain ordered down to low temperature. The large Mn moments tend not to align completely with the
ferromagnetic Co matrix, but rather maintain dynamically fluctuations below the Co moment ordering temperature
and even at high magnetic fields. At lower temperatures, the dynamic Mn moments freeze into a completely disordered re-entrant cluster spin glass
state while the Co sublattice remains mostly ferromagnetically ordered. This two-sublattice behavior allows for the coexistence
of disordered and spin-glass magnetic states with ordered magnetic states, such as helimagnetic, conical, and skyrmion lattices.

\section{Methods}
\subsection{Sample preparation}
\label{sec:preparation}
Five samples of Co$_x$Zn$_y$Mn$_z$ with nominal compositions \ce{Co10Zn10}, \ce{Co9Zn9Mn2}, \ce{Co8Zn10Mn2}, 
\ce{Co8Zn9Mn3}, and \ce{Co7Zn7Mn6} were prepared from the elements, using a procedure similar to previous reports \cite{Wei2016, Tokunaga2015}. Stoichiometric amounts of Co powder, 
Zn shot, and Mn pieces totaling 0.5\,g were weighed and sealed in an evacuated silica ampoule. The ampoule 
was heated to 1000\,$^\circ$C for 24\,hours and then slowly cooled to 925\,$^\circ$C over the course of 
75\,hours. Finally, the sample was held at 925\,$^\circ$C for 48 hours before quenching to room-temperature 
in a water bath. For \ce{Co8Zn9Mn3} and \ce{Co7Zn7Mn6} this same procedure was performed with 5\,g of starting 
material for neutron diffraction experiments. This solidification procedure yielded shiny metallic slugs. 
In cases where Mn was present, some green spots on the slug or green powder was found in the ampoule. X-ray 
diffraction revealed this green powder to be mainly MnO, and in all but the most Mn-rich sample (\ce{Co7Zn7Mn6}), 
this powder was easily removed from the metallic slug. The slugs were then pulverized for study by X-ray and 
neutron diffraction and magnetization.

\subsection{Diffraction and Rietveld Refinement}

After preparation, samples were initially checked for rough phase composition using a Pananalytical Empyrean X-ray diffractometer with Cu K$\alpha$ radiation ($\lambda$(K$\alpha$1)\,=\,1.54056\,\AA,  $\lambda$(K$\alpha$2)\,=\,1.54439\,\AA) equipped with a PIXcel 1D detector. Powdered samples were placed on a zero background plate and measured in Bragg-Brentano (reflection) geometry. In order to minimize the effects of Co and Mn fluorescence in the CuK$\alpha$ beam, the detector was set to reject low-energy photons. 

High resolution synchrotron powder diffraction data were collected on the \ce{Co10Zn10}, \ce{Co8Zn9Mn3}, and \ce{Co7Zn7Mn6} samples using beamline 11-BM at the Advanced Photon Source (APS), Argonne National Laboratory using an average wavelength of 0.414581 \AA. Patterns for \ce{Co8Zn9Mn3} were collected at 350\,K and 100\,K, for \ce{Co7Zn7Mn6} at 300\,K and 100\,K, and for \ce{Co10Zn10} at 300\,K.

Neutron powder diffraction data were collected on the \ce{Co8Zn9Mn3} and \ce{Co7Zn7Mn6} samples using the BT-1 32 detector neutron powder diffractometer at the NCNR, NIST. A Cu(311) monochromator with a 90$^\circ$ take-off angle, $\lambda$ = 1.5402(2) \AA, and in-pile collimation of 60 minutes of arc were used. Data were collected over the range of 3-168$^\circ$ 2$\theta$ with a step size of 0.05$^\circ$. Samples were sealed in vanadium containers of length 50\,mm and diameter 6\,mm inside a dry He-filled glovebox. A closed-cycle He cryofurnace was used for temperature control between 14\,K and 350\,K. Patterns were collected for \ce{Co8Zn9Mn3} at 350\,K, 100\,K, and 14\,K, and for \ce{Co7Zn7Mn6} at 300\,K, 150\,K, 100\,K, and 14\,K.

Rietveld refinement of all patterns was performed using the TOPAS Academic software. The synchrotron patterns were fit using full-Voight peaks with peak width determined by a standard crystallite size term and a microstrain term for the Lorentzian and Gaussian components (four parameters). In most cases, the Gaussian components of the Voight peaks refined to zero and purely Lorentzian peak shapes were used. Diffractometer peak asymmetry was handled using a fixed axial divergence asymmetry correction based on the instrument geometry. The neutron diffraction patterns were fit using purely Gaussian Stephens peak shapes \cite{Stephens1999} (two parameters) as well as a standard Gaussian size broadening term (one parameter). 

For \ce{Co8Zn9Mn3} and \ce{Co7Zn7Mn6}, synchrotron and neutron diffraction patterns were collected on identical samples at identical temperatures above and below the magnetic ordering 
temperatures. These pairs of patterns were simultaneously corefined to yield an optimal structure. In the corefinements, synchrotron and neutron phases were allowed to have different 
lattice parameters, peak shapes, and instrumental parameters. Atom positions, atomic displacement parameters, and weight percent of a MnO impurity in the 
\ce{Co7Zn7Mn6} sample were kept fixed between the two patterns. To limit possible systematic errors in the refinement results caused by the very large intensity of the synchrotron diffraction 
patterns compared to the neutron diffraction patterns, the weights of the synchrotron data points were globally decreased by a constant factor (0.2 for \ce{Co8Zn9Mn3} and 0.65 for 
\ce{Co7Zn7Mn6}). This factor was chosen so that $\sum_{i}{w_i(Y_{\rm{obs}}-Y_{\rm{calc}})^2} / N_{\rm{data}}$ was approximately equal for the synchrotron and neutron data sets.

The \ce{Co8Zn9Mn3} sample was analyzed for composition using X-ray fluorescence on a Rigaku Primus IV using semiquantitative analysis. A bulk piece of the sample was polished flat, and 
measured at 10 different 1mm spots across the surface of the sample to obtain an average composition and standard deviation. The room temperature densities of \ce{Co8Zn9Mn3} and 
\ce{Co7Zn7Mn6} powder samples were measured using a micromeritics AccuPyc II 1340 helium pycnometer. Ten measurements of the volume of each sample were performed and used to 
calculate an average and standard deviation of the density.

\subsection{Density Functional Theory}
In order to evaluate the relative energetics of atomic site preferences, density functional theory calculations were performed using the Vienna Ab initio Simulation Package (VASP) \cite{Kresse1996} using projector augmented wave (PAW) psuedopotentials \cite{Blochl1994, Kresse1999} within the Perdew-Burke-Ernzerhor (PBE) generalized gradient approximation (GGA) \cite{Perdew1996}.  Spin-orbit coupling was not included. Symmetrically distinct colorings of the disordered 20-atom cubic unit cell were generated using the CASM code \cite{VanderVen2010, Puchala2013}. The energies of each ordering were calculated by first performing two sequential structural optimizations (conjugate gradient algorithm), and then a static calculation with the relaxed structure. Energy convergence criteria of 10$^{-5}$\,eV for the electronic loops and 10$^{-4}$\,eV for the ionic loops were used. During the structural optimizations, unit cell shape, volume, and ion positions were allowed to relax. All atoms were initialized with an initial collinear magnetic moment of 3\,$\mu_B$. 

\subsection{Magnetic measurements}
Magnetic measurements were performed using a Quantum Design DynaCool PPMS equipped with a Vibrating Sample Magnetometer (VSM), using both the low-temperature mode (2K to 400 K) and the high-temperature oven option (300 K to 900 K). For the low-temperature mode, powdered samples were measured in polypropylene capsules. For the high-temperature mode, as-cast pieces were sanded into thin, flat pieces (1\,mg to 5\,mg) and cemented onto the high-temperature oven stick using alumina cement (Zircar). Various measurements at fixed fields and temperatures were performed for each sample.  
In order to probe the paramagnetic Curie-Weiss behavior, magnetization as a function of temperature under an applied field of 1\,T was collected upon warming above the magnetic transitions to temperatures up to 800\,K. At temperatures between 600\,K and 900\,K, each of the samples showed a kink in the magnetization that appeared irreversible and therefore was assumed to indicate some reaction with the sample environment. Therefore, data at and above this magnetic feature was discarded. Given these experimental constraints, linear Curie-Weiss fits with no residual magnetization term were obtained for all samples except \ce{Co10Zn10}. 

AC magnetic susceptibility was obtained on the \ce{Co7Zn7Mn6} sample using a Quantum Design MPMS XL. The real ($\chi'$) and imaginary ($\chi''$) parts of the susceptility as a function of temperature was measured on the sample of \ce{Co7Zn7Mn6} under a DC field $H$\,=\,20\,mT with various excitation frequencies. The sample was measured upon warming, after cooling under the 20\,mT DC field. 

In addition, indirect magnetocaloric measurements were performed using DC magnetization. The magnetic entropy change associated with isothermal magnetization of a material $\Delta S_M$, which is a function of the temperature $T$ and applied magnetic field $H$, is obtained from magnetization data using the Maxwell relation:
\begin{equation}
\label{eqn:maxwell}
\left(\frac{\partial S}{\partial H}\right)_T =\left(\frac{\partial M}{\partial T}\right)_H
\end{equation}

\noindent In this equation, $S$ is the total entropy, $H$ is the magnetic field, $M$ is the magnetization, and $T$ is the temperature. Therefore, the isothermal entropy change upon application of field $H$ can be calculated from magnetic measurements at many fields and temperatures  by:

\begin{equation} 
\label{eqn:deltaSm}
\Delta S_M(H, T) = \int_{0}^{H}{\left(\frac{\partial M}{\partial T}\right)_{H'} dH'}
\end{equation}

\noindent The underlying magnetization data for this calculation was collected as continuously collected magnetization \emph{vs.} temperature sweeps under several different fields (Supplemental Material Fig.\,S3). The derivatives were calculated using a regularization procedure \cite{Stickel2010}, as described in Ref.~\onlinecite{Bocarsly2018}, and the integrals were calculated using the trapezoid method.

\section{Results and discussion}

\subsection{Atomic structure and site preferences}
\begin{figure*}[htb!]
	\centering
	\includegraphics[width=\textwidth]{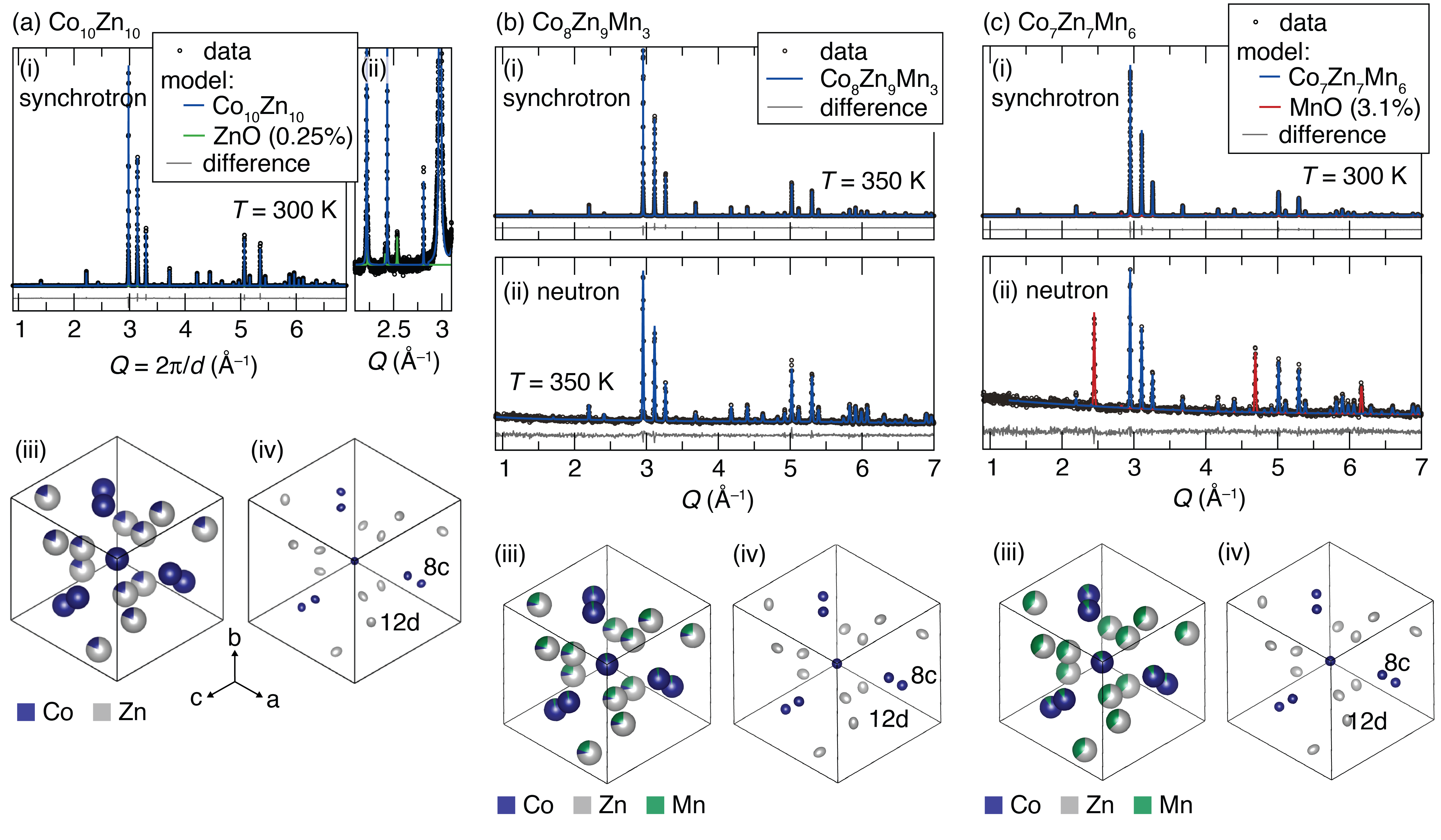}
	\caption{Synchrotron and paramagnetic neutron diffraction patterns for (a) \ce{Co10Zn10}, (b) \ce{Co8Zn9Mn3}, and (c) \ce{Co7Zn7Mn6}. For \ce{Co10Zn10} (a), the synchrotron diffraction pattern is shown along with Rietveld fit to the $P4_132$ $\beta$-Mn structure (i). A 0.25\,wt.\% ZnO secondary phase is observed, which can be seen in an expanded view (ii). Refined structures are shown both with partial occupancies indicated by the atom coloring (iii) and anisotropic atomic displacement parameters displayed as 90\% probability ellipsoids (iv). For \ce{Co8Zn9Mn3} (b) and \ce{Co7Zn7Mn6} (c), combined synchrotron (i) and neutron (ii) Rietveld refinements above the magnetic Curie temperatures are shown. The \ce{Co8Zn9Mn3} shows only the $\beta$-Mn phase, while the \ce{Co7Zn7Mn6} sample shows a 3.12(3)\,wt.\% MnO secondary phase, which, due to structure factor differences between X-ray and neutron diffraction, appears much stronger in the neutron pattern. The refined structural details are described in table~\ref{tbl:rietveld}~and~\ref{tbl:site-properties}.}
	\label{fig:all-structs}
\end{figure*}

Rietveld refinement of synchrotron powder diffraction for \ce{Co10Zn10}  is shown in Fig.\,\ref{fig:all-structs}a., the reported $\beta$-Mn structure ($P4_132$) \cite{Xie2013} fits the observed synchrotron diffraction pattern at room temperature very well, with a small wurtzite ZnO \cite{Albertsson1989} impurity (0.25(2) wt\%). The 8c site is found to be uniformly occupied by Co, while the 12d site has a random distribution of about 80\% Zn and 20\% Co, giving a final composition of Co$_{10.24(14)}$Zn$_{9.76(14)}$, very close to the nominal starting composition of \ce{Co10Zn10}. No peaks violating this model are found, and the data quality is sufficient to refine well-behaved anisotropic displacement parameters (Fig.\,\ref{fig:all-structs}a-iv). The structural parameters determined using this refinement can be found in tables~\ref{tbl:rietveld}~and~\ref{tbl:site-properties}. 

As Mn is added to this structure, the site preferences become more complex to determine. X-ray diffraction, which is sensitive to the electron density of a crystal structure, is poorly suited to distinguishing between elements with similar electron count, such as Mn and Co. On the other hand, neutrons are scattered by nuclear interactions and magnetization density, and can therefore give complementary information to X-ray diffraction. At first glance, the large contrast between the neutron scattering lengths of Co, Zn and Mn (2.48\,fm, 5.68\,fm and $-$3.73\,fm, respectively) would seem to make neutron powder diffraction an ideal tool to study the atomic 
site preferences in these materials. However, since there are only two sites in the $\beta$-Mn structure, only a single compositional degree of freedom may be stably refined using the Rietveld method (other degrees 
of freedom will correlate with the pattern scale factor). In other words, when performing Rietveld refinement, one cannot allow occupancies for Co, Zn, and 
Mn on the two sites to refine independently, but must constrain the compositions such that only a single compositional parameter is refined.
Simultaneous refinements between synchrotron XRD and neutron diffraction can improve the overall fit and allow for refinement of an additional compositional degree of freedom, although 
the compositional information given by the synchrotron pattern is fairly limited. Therefore, in order to determine 
the correct ordering in Co$_x$Zn$_y$Mn$_z$ compounds, we jointly refine the synchrotron and neutron data using several plausible models, and distinguish between equally-well-fitting models by comparing their predicted results to reference measurements such as X-ray Fluorescence compositional analysis and density measurements performed using helium gas pycnometry.

Figure\,\ref{fig:all-structs}b shows neutron and synchrotron diffraction patterns taken on a single sample of material with nominal composition 
\ce{Co8Zn9Mn3}. A joint Rietveld refinement between the two histograms has been performed. Both datasets shows no peaks except those expected for 
the $\beta$-Mn phase, indicating that the sample is phase pure. Rietveld refinement was performed using several compositional models. Given that the 8c site is entirely filled by Co in Co-Zn compounds \cite{Wei2016},
 a natural starting 
guess is a configuration where the 8 Co atoms per unit cell sit on the 8c site and 9 Zn atoms and 3 Mn atoms are randomly distributed on the 12d site. This gives a reasonably good fit to the data (overall $R_{wp}
$=12.38). If Mn has a strong site-preference for the 8c site, on the other hand, it would be expected to displace Co from the 8c site to the 12d site, which gives far worse fit ($R_{wp}$=13.78). In addition, various models with Zn on the 8c all give very 
poor fits. 

The fit may be improved by allowing a fraction of the Mn (14\%) to move to the 8c site, displacing an equivalent amount of Co to the 12d site ($R_{wp}$\,=\,12.36). No such improvement is found when allowing Zn to move 
onto the 8c site. This model fits the synchrotron and neutron data very well and gives physically reasonable refined structural parameters with the overall composition \ce{Co8Zn9Mn3}. However, we may expect that the sample deviates from the 
nominal composition due to the observation that MnO was formed and discarded during the sample preparation. Therefore, we analyzed the sample using X-ray fluorescence (XRF) compositional analysis, which gave an Mn-deficient composition, 
Co$_{8.302(14)}$Zn$_{8.869(15)}$Mn$_{2.829(6)}$. Rietveld refinements assuming this empirical composition were also performed, and yielded very similar results to the nominal composition refinements, with a slightly better overall fit ($R_{wp}$\,=\,12.35). In this model, 18\% of the Mn is found on the 8c site while  82\% is on the 12d site, giving a site configuration of (Co$_{7.58(3)}$Mn$_{0.42(3)}$)$^{\rm{8c}}$(Zn$_{8.87}$Mn$_{2.41(3)}$Co$_{0.72(3)}$)$^{\rm{12d}}$. The refined density 
(7.805(7)\,g\,cm$^{-3}$) is in agreement with the density determined by helium pycnometry (7.78(5)\,g\,cm$^{-3}$). The fit given by this model, as well as drawings of the determined structure, is shown in Fig.\,\ref{fig:all-structs}b. In addition, refined 
parameters are reported in tables~\ref{tbl:rietveld}~and~\ref{tbl:site-properties}.

We also performed synchrotron and neutron corefinements for data collected on a \ce{Co7Zn7Mn6} sample, which is a relatively Mn-rich 
Co$_x$Zn$_y$Mn$_z$ composition (Fig.\,\ref{fig:all-structs}c). This sample was found to contain about 3\% MnO by weight, present in equal quantities in the neutron diffraction pattern and the synchrotron diffraction pattern. Fortuitously, this impurity serves as an internal standard between the two scans, fixing the scale factor of the neutron pattern relative to the synchrotron pattern. This constraint allows for stable refinement of an additional compositional degree of freedom, and, as a 
result, there is much less ambiguity in this refinement than in the refinement of \ce{Co8Zn9Mn3}. Based on the structures of the \ce{Co10Zn10} and 
\ce{Co8Zn9Mn3} samples, a natural first guess for the structure of \ce{Co7Zn7Mn6} is a model with Co and Mn sitting on the 8c site, and Mn and Zn on the 
12d site. 

\begin{table*}[]
\centering
\caption{Composition, phase analysis, and lattice parameters from Rietveld refinement of  synchrotron and neutron diffraction of Co$_x$Zn$_y$Mn$_z$ samples. Samples are refined in the $P4_132$ spacegroup (No. 213). Numbers in parentheses are standard uncertainties in the last given digit(s) from Rietveld refinement. For site-specific properties and magnetic moments refined from these datasets, see table~\ref{tbl:site-properties} and table~\ref{tbl:mag-refinements}.}
\begin{tabular}{@{\extracolsep{4pt}}ll@{\hspace{0.3cm}}cclrrr}
\toprule
composition                                                                     & secondary phases & $T$   & source\footnotemark[1] & \multicolumn{1}{c}{$a$}   & $R_{\rm{wp}}$ & $R_{\rm{p}}$ & $R_{\rm{exp}}$ \\
nominal (determined)                                                            &                  & (K)   &                        & \multicolumn{1}{c}{(\AA)} &               &              & \\ \colrule
\ce{Co10Zn10}                                                                   & ZnO (0.25(2)\%)  & 300 & S                      & 6.322506(3)               & 15.58         & 12.18        & 9.37  \\
\hspace{0.3cm} (Co$_{10.24(14)}$Zn$_{9.76(14)}$)\footnotemark[2] & \\[3ex] 
\ce{Co8Zn9Mn3}                                                                  & none observed    & 350 & S                      & 6.380518(6)               & 13.44         & 10.32        & 6.43  \\
\hspace{0.3cm}(Co$_{8.302(14)}$Zn$_{8.869(15)}$Mn$_{2.829(6)}$)\footnotemark[3] &                  &       & N                      & 6.3787(1)                 & 6.09          & 4.96         & 6.65  \\
                                                                                &                  & 100 & S                      & 6.359298(4)               & 11.31         & 8.59         & 6.32  \\
                                                                                &                  &       & N                      & 6.3862(1)                 & 6.39          & 5.06         & 6.47  \\
                                                                                &                  & 14  & N                      & 6.3558(1)                 & 6.52          & 5.29         & 6.60  \\[3ex]
\ce{Co7Zn7Mn6}                                                                  & MnO (3.12(3)\%)  & 300 & S                      & 6.39108(1)                & 9.84          & 7.96         & 8.69  \\
\hspace{0.3cm}(Co$_{7.04(4)}$Zn$_{7.43(3)}$Mn$_{5.53(5)}$)\footnotemark[2]      &                  &       & N                      & 6.3889(2)                 & 4.27          & 3.52         & 4.80  \\
                                                                                &                  & 150 & N                      & 6.3745(2)                 & 5.91          & 4.79         & 6.71  \\
                                                                                &                  & 100 & S                      & 6.37253(1)                & 9.62          & 8.19         & 10.16 \\
                                                                                &                  &       & N                      & 6.3705(1)                 & 4.44          & 3.67         & 4.81  \\
                                                                                &                  & 14  & N                      & 6.3692(1)                 & 3.52          & 2.90         & 3.50  \\[3ex]
\ce{Co9Zn9Mn2}                                                                  & none observed    & 300 & L                      & 6.35441(9)                & 3.68          & 2.87         & 3.33  \\[3ex]
\ce{Co8Zn10Mn2} (Co$_{8.8}$Zn$_{8.7}$Mn$_{2.5}$)\footnotemark[4] & Co$_{2.6}$Zn$_{10.0}$ (20.1(6)\%) & 300               & L                      & 6.3645(2)                 & 4.14          & 3.23         & 3.75  \\[3ex]
\botrule
\footnotetext[1]{synchrotron (S), laboratory X-ray (L), or neutron (N)}
\footnotetext[2]{determined by Rietveld refinement}
\footnotetext[3]{determined by x-ray fluorescence}
\footnotetext[4]{determined by phase balance with nominal composition}
\end{tabular}

\label{tbl:rietveld}
\end{table*}

\begin{table*}[]
\caption{Site-specific structural properties of Co$_x$Zn$_y$Mn$_z$ samples from Rietveld refinement of synchrotron and neutron diffraction. The 8c site has coordinates ($x$,$x$,$x$), and the 12d site has coordinates ($\frac18$,$y$,$y+\frac14$). Numbers in parentheses are standard uncertainties in the last given digit(s) from Rietveld refinement.}
\begin{tabular}{@{\extracolsep{5pt}}llclrrlrr}
\toprule
    & $T$ (K) & refinement & \multicolumn{3}{c}{8c}                                      & \multicolumn{3}{c}{12d}                                                \\ \cline{4-6} \cline{7-9}
               &    & type\footnotemark[1]   & composition                  & \multicolumn{1}{c}{$x$}          & \multicolumn{1}{c}{$B_{\rm{eq}}$} & composition                             & \multicolumn{1}{c}{$y$} & \multicolumn{1}{c}{$B_{\rm{eq}}$} \\ \colrule
 \ce{Co10Zn10} & 300  & S & Co$_8$                       & 0.06429(5) & 0.419(9) & Co$_{2.24(15)}$Zn$_{9.76(15)}$          & 0.20296(4) & 0.588(7) \\[1ex]
\ce{Co8Zn9Mn3} & 350  & S+N       & Co$_{7.58(3)}$Mn$_{0.42(3)}$ & 0.06479(4) & 0.670(8) & Co$_{0.72(3)}$Zn$_{8.87}$Mn$_{2.41(3)}$ & 0.20279(3) & 0.812(6) \\
               & 100  & S+N       &                              & 0.06522(3) & 0.269(5) &                                         & 0.20299(2) & 0.345(4) \\
               & 14   & N     &                              & 0.0652(4)  & 0.25(5)  &                                         & 0.2029(2)  & 0.27(4)  \\[1ex]
\ce{Co7Zn7Mn6} & 300  & S+N       & Co$_{7.04(4)}$Mn$_{0.96(4)}$ & 0.06477(3) & 0.521(5) & Zn$_{7.43(3)}$Mn$_{4.57(3)}$            & 0.20271(2) & 0.703(5) \\
               & 150  & N     &                              & 0.0637(7)  & 0.6(1)   &                                         & 0.2029(4)  & 0.30(6)  \\
               & 100  & S+N       &                              & 0.06489(2) & 0.100(4) &                                         & 0.20270(2) & 0.190(3) \\
               & 14   & N     &                              & 0.0649(4)  & 0.1(1)   &                                         & 0.2029(2)  & 0.17(4)  \\ \botrule\botrule
\footnotetext[1]{synchrotron (S), neutron (N), or joint synchrotron and neutron (S+N)}
\end{tabular}
\label{tbl:site-properties}
\end{table*}

The compositions of both sites may be independently refined, yielding a structure that matches both the synchrotron and neutron patterns very well, as 
shown in Fig.\,\ref{fig:all-structs}c and reported in tables~\ref{tbl:rietveld}~and~\ref{tbl:site-properties}. This refinement gives a total composition of 
Co$_{7.04(4)}$Zn$_{7.43(3)}$Mn$_{5.53(5)}$, with a 3.12(3)\% MnO secondary phase. The mixture of these phases (neglecting oxygen) gives a total 
composition of Co$_{6.85(2)}$Zn$_{7.23(2)}$Mn$_{5.92(4)}$, very close to the nominal composition. In addition, the mixture density is calculated to be 
7.56(2)\,g\,cm$^{-3}$, in agreement with a measurement of the density determined by He pycnometry (7.53(2)\,g\,cm$^{-3}$).

Samples of \ce{Co9Zn9Mn2} and \ce{Co8Zn10Mn2} were also prepared and checked for phase content with laboratory XRD. The patterns are found in Supplemental Material Fig.\,S3. The \ce{Co9Zn9Mn2} sample is found to be pure $\beta$-Mn,
with no observable secondary phases at the resolution of the experiment. \ce{Co8Zn10Mn2} was found to contain about 20\% of $\gamma$-brass Co$_{2.6}$Zn$_{10}$, which is observed in Co-Zn materials when the Zn:Co ratio deviates too far 
above 1:1 \cite{Xie2014}. By mass balance, the remaining 80\% $\beta$-Mn phase is expected to have the composition Co$_{8.8}$Zn$_{8.7}$Mn$_{2.5}$, which has a nearly 1:1 Zn:Co ratio. The $\gamma$-brass has a very low magnetic moment and therefore is not 
expected to have a significant impact on bulk magnetization measurements, and so this sample was included in the magnetization studies with the mass corrected to remove the nonmagnetic impurity.

Although all samples exhibit slight deviations from the nominal compositions, the samples will be referred to by their nominal compositions for the remainder of this article. The experimental compositions are used in calculating moments and in
performing magnetic neutron refinements. Crystallographic information files containing the final 
refined structures, including anisotropic atomic displacement parameters, for the \ce{Co10Zn10}, \ce{Co8Zn8Mn3}, and \ce{Co7Zn7Mn6} samples are included in the Supplemental Material.

\subsection{Energetics of site preferences}

\begin{table}[!bh]
\centering
\caption{DFT calculations of \ce{Co10Zn10} with various atomic configurations. For each configuration, five unit cells with randomly selected orderings of Co and Zn have been calculated, and average properties (with standard deviations) are reported. Energies are given in reference to the energy of \ce{Co8}-\ce{Co2Zn10}.}

\label{tbl:dft_1010}
\begin{tabular}{@{\extracolsep{2pt}}lcrrrr}
\toprule
 configuration     & Zn on  & \multicolumn{1}{c}{$\Delta E$} & \multicolumn{2}{l}{Co moment ($\mu_B$)} & volume \\  \cline{4-5}
 ([8c]-[12d]) &  8c site    & (meV) & \multicolumn{1}{c}{8c}  & \multicolumn{1}{c}{12d}    & \multicolumn{1}{c}{(\AA$^3$)} \\ \colrule
\ce{Co8}-\ce{Co2Zn10}   & 0 & 0(114)   & 1.33(7)  & 1.2(3)  & 248.0(6) \\ 
\ce{Co7Zn1}-\ce{Co3Zn9} & 1 & 596(64)  & 1.32(12) & 1.43(7) & 249.1(3) \\ 
\ce{Co6Zn2}-\ce{Co4Zn8} & 2 & 1067(69) & 1.31(15) & 1.48(7) & 249.8(5) \\ 
\ce{Co5Zn3}-\ce{Co5Zn7} & 3 & 1564(88) & 1.29(11) & 1.47(7) & 251.0(4) \\ \botrule            
\end{tabular}
\end{table}

\begin{table*}[htb]
\centering
\caption{Results of spin-polarized DFT calculations of \ce{Co8Zn9Mn3} with various atomic configurations. For each configuration, five unit cells with randomly selected orderings of Co, Zn, and Mn have been calculated, and average properties (with standard deviations) are reported. The final column refers to non-spin-polarized DFT calculations on the same unit cells. Energies are provided relative to the \ce{Co_8}-\ce{Zn9Mn3} calculations from either spin-polarized (column 2) or non-spin-polarized (final column) calculations. The non-spin polarized reference is 2.518 eV higher in energy than the spin-polarized reference.}

\label{tbl:dft_893}
\begin{tabular}{@{\extracolsep{6pt}}lcrrrrrrrrr}
\toprule
      configuration           & non-Co atoms  & \multicolumn{1}{c}{$\Delta E$} & \multicolumn{2}{l}{Mn moment ($\mu_B$)} & \multicolumn{2}{l}{Co moment ($\mu_B$)} & volume      & nonmag. $\Delta E$ \\ \cline{4-5} \cline{6-7} 
    ([8c]-[12d]) & on 8c site    & (meV)      & \multicolumn{1}{c}{8c}   & \multicolumn{1}{c}{12d}   & \multicolumn{1}{c}{8c} & \multicolumn{1}{c}{12d}   & \multicolumn{1}{c}{(\AA$^3$)}  & \multicolumn{1}{c}{(meV)}    \\ \colrule
\ce{Co8}-\ce{Zn9Mn3}       & 0             & 0 (63)   & \multicolumn{1}{c}{-} & 3.28(5)               & 1.39(4)  & \multicolumn{1}{c}{-} & 253.5(2) & 0(359)\\          
                           & Zn on 8c site &          &                       &                       &          &                       &          &          \\          
\ce{Co7Zn1}-\ce{Co1Zn8Mn3} & 1             & 676(118) & \multicolumn{1}{c}{-} & 3.23(4)               & 1.33(5)  & 1.31(9)               & 253.5(3) & 216(139) \\
\ce{Co6Zn2}-\ce{Co2Zn7Mn3} & 2             & 1260(54) & \multicolumn{1}{c}{-} & 3.21(7)               & 1.21(14) & 1.26(15)              & 253.5(4) & 555(114) \\
\ce{Co5Zn3}-\ce{Co3Zn6Mn3} & 3             &  1595(99)        & \multicolumn{1}{c}{-} & 3.24(8)               & 1.16(18) & 1.39(7)               & 254.0(7) & 1218(308) \\
                           & Mn on 8c site &          &                       &                       &          &                       &          &          \\          
\ce{Co7Mn1}-\ce{Co1Zn9Mn2} & 1             & 182(51)  & 2.39(8)               & 3.26(6)               & 1.23(12) & 1.25(5)               & 251.1(2) & $-$67(197) \\         
\ce{Co6Mn2}-\ce{Co2Zn9Mn1} & 2             & 441(113) & 2.1(7)                & 3.18(6)               & 1.13(19) & 1.35(9)               & 249(1)   & $-$648(183) \\        
\ce{Co5Mn3}-\ce{Co3Zn9}    & 3             & 628(134) & 2.40(11)              & \multicolumn{1}{c}{-} & 1.0(3)   & 1.39(8)               & 249.2(7) & $-$1014(159) \\ \botrule 
\end{tabular}
\end{table*}

In order to verify and rationalize the observed patterns of site preferences and site mixing in Co$_x$Zn$_y$Mn$_z$, we performed density functional theory 
(DFT) calculations. Table~\ref{tbl:dft_1010} and Table~\ref{tbl:dft_893} show the result of DFT calculations for various hypothetical unit cells with the overall compositions \ce{Co10Zn10} and \ce{Co8Zn9Mn3}, respectively. In each case, we start with a presumed ``ground state'' for each composition with Co completely occupying the 8c site and the remaining elements randomly occupying the 12d site. Then, we consider alternate configurations generated by taking between one and three Zn or Mn atoms from the 12d site and swapping them with Co atoms from the 8c site. Because all of the configurations studied, including the ``ground state'', involve atomic site disorder, five randomly selected colorings of the single 20-atom $\beta$-Mn unit cell were considered for each configuration, with the average energy (with standard deviation) of those five cells reported.

As can be seen in tables\,\ref{tbl:dft_1010}~and~\ref{tbl:dft_893}, the energy of the unit cell increases monotonically as Zn is switched onto the 8c site in \ce{Co10Zn10} and as Mn or Zn is switched onto the 8c site in \ce{Co8Zn9Mn3}. Linear regression of the calculated cell energies \emph{vs.} the number of non-Co atoms on the 8c atoms gives a rough estimate of the energy penalty associated with these sorts of anti-site defects. In \ce{Co10Zn10}, this penalty is 516(20) meV per Zn atom moved onto the 8c site. In \ce{Co8Zn9Mn3}, the penalty is 210(20) meV/Mn atom on the 8c site and 540(30) meV/Zn atom on the 8c site. 

Our experimental samples were prepared by heating to high temperature, slowly cooling to 925$^\circ$C\,=\,103 meV/$k_B$ and then quenching to room temperature. Therefore, we expect the room temperature samples to contain some amount of disorder frozen in during the quench from high temperature. However, based on the quench temperature (103\,meV/$k_B$), we expect that Zn atoms swapped onto the 8c site ($\Delta E$ = 516 meV to 540 meV) will be quite rare in both \ce{Co10Zn10} and \ce{Co8Zn9Mn3}. In \ce{Co8Zn9Mn3}, however, Mn may be swapped onto the 8c site at a far lower energy penalty (210\,meV\,$\approx 2k_BT$) and therefore we expect a substantial amount of Mn to be found on the 8c site in the samples prepared by quenching, even though this disordered configuration represents a non-minimum internal energy. This prediction matches our refined structure for \ce{Co8Zn9Mn3} exactly, which has Mn predominantly on the 12d site, with some mixing (0.43(3) Mn per 20 atom unit cell) onto the 8c site. Based on this energetic analysis, we expect that that quantity of Mn on the 8c site could be reduced by slowly cooling to room temperature (300 K = 26 meV) rather than quenching from high temperature, although we have not explored this experimentally.

In addition to validating the Rietveld refinements, the DFT calculations also allow us to probe the origin of the site preferences exhibited by Co, Zn, and Mn. According to Xie \emph{et al.} \cite{Xie2013}, the site preference of Co
for the 8c site in CoZn may be explained by the smaller size of metallic Co compared to Zn (125\, pm and 134\,pm 12-coordinate atomic radii \cite{Kaye1995}.) Indeed, tables~\ref{tbl:dft_1010}~and~\ref{tbl:dft_893} show the unit cell volume is seen to increase as Zn is switched to the 8c site in both \ce{Co10Zn10} and \ce{Co8Zn9Mn3}, indicating that configurations with Zn on the 8c have non-optimal packing, due to a size mismatch.

The case of Mn, on the other hand, is more complex. Mn has 12-coordinate atomic radius of 127 pm, and therefore should behave similarly to Co on the basis of size arguments. However, the DFT calculations show that unit cell volume actually decreases as Mn is switched to the 8c site, despite the energy of the cell increasing. This suggests that atomic packing considerations favor Mn on the 8c site, even more strongly than Co. However, the calculations reveal a key difference between Co and Mn: magnetism. In both \ce{Co10Zn10} and \ce{Co8Zn9Mn3}, each Co atom generally holds a local moment of between 1.1 and 1.4 $\mu_B$, with no clear systematic difference between the 8c and 12d site. Mn, on the other hand, holds a moment of between 3.2 and 3.3 $\mu_B$ if it is on the 12d site and a moment between 2.1 and 2.4 $\mu_B$ if it is on the 8c site (Table~\ref{tbl:dft_893}).

The larger Mn moment observed when the Mn is on the 12d site makes sense, as the larger site allows the Mn $d$-orbitals to localize more than the smaller 8c site. The formation of a large local moment on Mn provides a magnetic energy savings for the structure, which is seen in the difference in the non-spin-polarized DFT energies (last column of Table~\ref{tbl:dft_893}) and spin-polarized DFT energies for the two structures. As Mn is moved to the 8c site, and its moment decreases, the stabilization due to spin-polarization of the electronic structure decreases. In fact, the non-spin-polarized calculations show that, in the absence of magnetism, Mn prefers to sit on the 8c site, displacing Co to the 12d site. Therefore, we may conclude that the Mn site-preferences are driven by a competition between size and magnetism: the non-optimal packing associated with having Mn on the 12d site is offset by the energy savings associated with the development of large local moment on the Mn atoms when they are on the 12d site.

In the case of \ce{Co7Zn7Mn6}, there is not enough Co to fill the 8c site and so the lowest energy state is expected to contain 7/8 Co and 1/8 Mn on the 8c site. DFT calculations of this configuration (\ce{Co7Mn1}-\ce{Zn7Mn5}) give Mn moments of 3.18(7)\,$\mu_B$ and 1.9(1.0)\,$\mu_B$ for the 12d and 8c sites, respectively and Co moments of 1.30(14)\,$\mu_B$ for the 8c site. These moments are very similar to the moments calculated for \ce{Co10Zn10} and \ce{Co8Zn9Mn3}, suggesting that the atoms, and in particular Mn on the 12d sites, hold largely localized magnetic moments. While a full energetic analysis of the site preferences of this highly disordered composition has not been performed, we expect the energy cost of switching additional Mn or Zn atoms from the 12d site to the 8c site in \ce{Co7Zn7Mn6} to be similar to those calculated for \ce{Co10Zn10} and \ce{Co8Zn9Mn3}. However, because the 8c site is already substantially disordered in \ce{Co7Zn7Mn6}, switching additional Mn onto that site will yield only a modest increase in entropy compared to switching Mn onto the completely ordered 8c site in \ce{Co8Zn9Mn3}. This explains why we do not observe a significant amount of Co on the 12d site in the \ce{Co7Zn7Mn6} sample.

Based on neutron and synchrotron powder diffraction data, compositional probes, and density functional theory calculations, we can build up a picture of how the 8c and 12d sites are expected to 
be populated in arbitrary Co$_x$Zn$_y$Mn$_z$ compounds. The preference for the 8c site is as follows: Co $>$ Mn $\gg$ Zn. Therefore, given a Co$_x$Zn$_y$Mn$_z$ composition, up to 8 Co 
atoms will fill the 8c site, with the remainder (if $x<$\,8) being filled by Mn. The remaining Co, Zn, and Mn atoms will be found on the 12d site. If $x \ge 8$, and there is Mn present, a fraction of the 
Mn is likely to mix onto the 8c site, particularly in samples quenched from high temperature. This means that, in all samples containing Mn, there is substantial compositional disorder on both 
atomic sites, although the 12d site is generally more disordered. While atomic packing considerations would favor Mn on the 8c site, it instead favors the larger 12d site because of its ability to 
develop a large (approximate 3\,$\mu_B$) local moment on that site.

\subsection{Magnetic Properties}

\begin{figure*}[ht!]
	\centering
	\includegraphics[width=\textwidth]{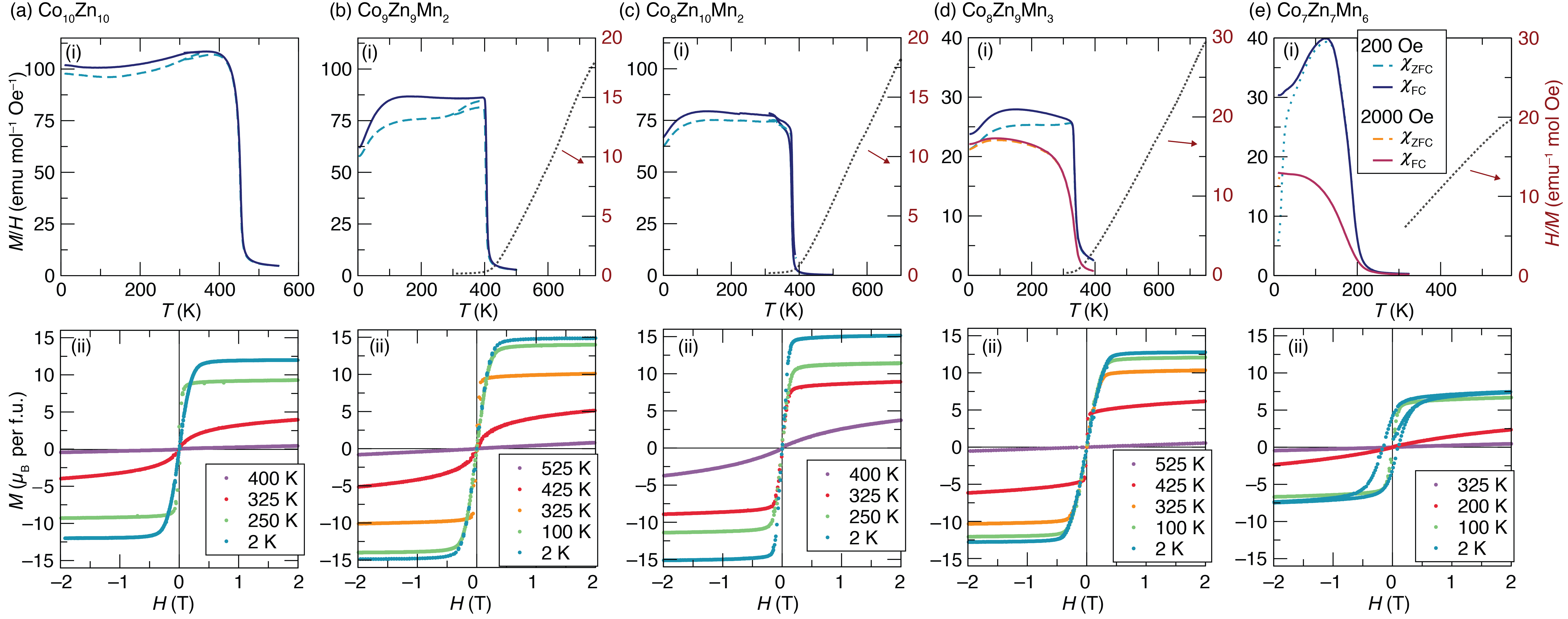}
	\caption{Magnetization data for Co$_x$Zn$_y$Mn$_z$ samples. The top row shows susceptibility $M/H$ as a function of temperature under an applied field $H$\,=\,20\,mT, showing both zero-field cooled (ZFC, dashed dashed-lines) and field-cooled (FC, solid lines) measurements. For \ce{Co8Zn9Mn3} (d) and \ce{Co7Zn7Mn6} (e), measurements taken at $H$\,=\,200\,mT, are also shown. For all samples except \ce{Co10Zn10}, the high-temperature magnetization, taken under an applied field $H$\,=\,1\,T is shown as inverse susceptibility ($\chi^{-1}$, dotted lines), demonstrating Curie-Weiss behavior at high temperature. The bottom row shows magnetization $M$ as a function of applied field $H$ at various temperatures for each material. For each temperature, a full five-branch hysteresis loop is shown, although no significant magnetic hysteresis is observed in any loop except for at 2\,K in \ce{Co7Zn7Mn6}. The unit emu\,mol$^{-1}$\,Oe$^{-1}$ is equal to 4$\pi \times 10^{-6}$\,m$^3$.
	}
	\label{fig:mag-data}
\end{figure*}

\begin{table*}[hbt]
\centering
\caption{Magnetic properties for Co$_x$Zn$_y$Mn$_z$ samples. $a$ refers to the lattice parameter, $T_C$ is the onset temperature of magnetic ordering, $\theta_{\rm{CW}}$ and $\mu_{\rm{eff, CW}}$ refer to the results of Curie-Weiss fits of high-temperature magnetic susceptibility taken at $H$\,=1\,T, and $M_{\rm{sat}}$ refers to saturation magnetization at low temperature, expressed either in gravimetric units or in Bohr magnetons per formula unit or per magnetic ion (Mn and Co). $\Delta S_M^{\rm{pk}}$ is the peak value of the entropy change upon isothermal magnetization to $H$\,=\,2\,T or $H$\,=\,5\,T.
}
\label{tbl:mag}
\begin{tabular}{@{\extracolsep{3pt}}lccccccccc}

\toprule
            &   $a$   &  $T_C$   & $\theta_{\rm{CW}}$ & \multicolumn{3}{c}{$M_{\rm{sat}}$, $T$ = 2\,K, $H$ = 5\,T}       &     $\mu_{\rm{eff}}$   & \multicolumn{2}{c}{$\Delta  S_{M}^{\rm{pk}}$ (J kg$^{-1}$ K$^{-1}$)} \\  \cline{5-7} \cline{9-10}
composition & (\AA)  &  (K)  & (K) & (Am$^2$kg$^{-1}$) & ($\mu_{\rm{B}}$/f.u) & ($\mu_{\rm{B}}$/mag. ion)  &  ($\mu_{\rm{B}}$/f.u.) & $H$ = 2\,T  & $H$ = 5\,T           \\ \colrule
\ce{Co10Zn10}      & 6.3226\footnotemark[1] & 470 &    - -\footnotemark[3]  & 54.0  & 12.0   & 1.20   &    - -\footnotemark[3] &  --0.86 &  --1.58  \\
\ce{Co9Zn9Mn2}   & 6.3544\footnotemark[2] & 415 & 432(2)     &  67.9  & 14.9  &  1.35  &  11.9  &  --1.06  &  --1.94  \\
\ce{Co8Zn10Mn2} & 6.3645\footnotemark[2] & 395 & 395.5(4)  &  70.0  & 15.3  &  1.53  &  11.5  &  --1.16  &  --2.20  \\
\ce{Co8Zn9Mn3}   & 6.3806\footnotemark[1] & 340 & 363.7(9)  &  58.8  & 12.9  &  1.17  &  10.3  &  --0.94  &  --1.79 \\
\ce{Co7Zn7Mn6}   & 6.3911\footnotemark[1] & 210 & 183.1(9)  &  33.5  &  7.2   &  0.55  &  11.9  &  --0.46  &  --0.97  \\ \botrule
\footnotetext[1]{synchrotron XRD} \footnotetext[2]{laboratory XRD} \footnotetext[3]{not obtained due to experimental complications}
\end{tabular}
\end{table*}

\begin{figure}[t!]
	\centering
	\includegraphics[width=\columnwidth]{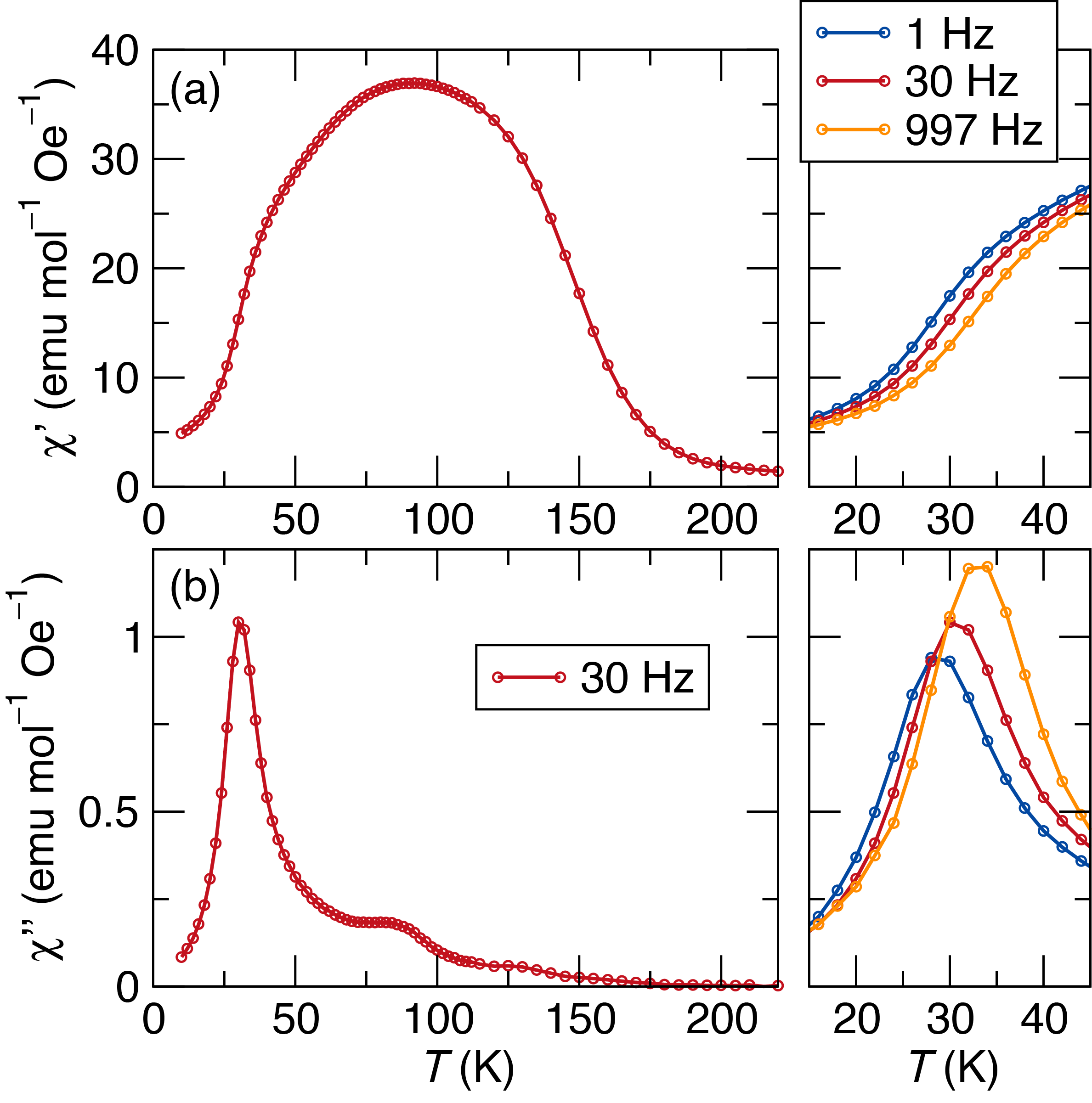}
	\caption{AC magnetic susceptibility measurements of \ce{Co7Zn7Mn6}. (a) Shows the real (in-phase, $\chi'$) part of the AC susceptibility, while (b) shows the imaginary (out-of-phase, $\chi''$) part. At high temperatures, the AC susceptibility resembles the DC susceptibility shown in Fig.\,\ref{fig:mag-data}. However, at low temperatures, $\chi'$ drops and there is a peak in $\chi''$ indicating glassy dynamics of the spins. The right side of the figure shows the  dependence of this feature on the frequency of the applied AC field. The locations of the peak in $\chi''$ increases by about 4.5\,K as the excitation field frequency is increased from 1\,Hz to 997\,Hz. The measurements were collected while warming, after cooling under a DC field $H$\,=\,20\,mT. The unit emu\,mol$^{-1}$\,Oe$^{-1}$ is equal to 4$\pi \times 10^{-6}$\,m$^3$.
	}
	\label{fig:acms}
\end{figure}

\begin{figure}[t!]
	\centering
	\includegraphics[width=\columnwidth]{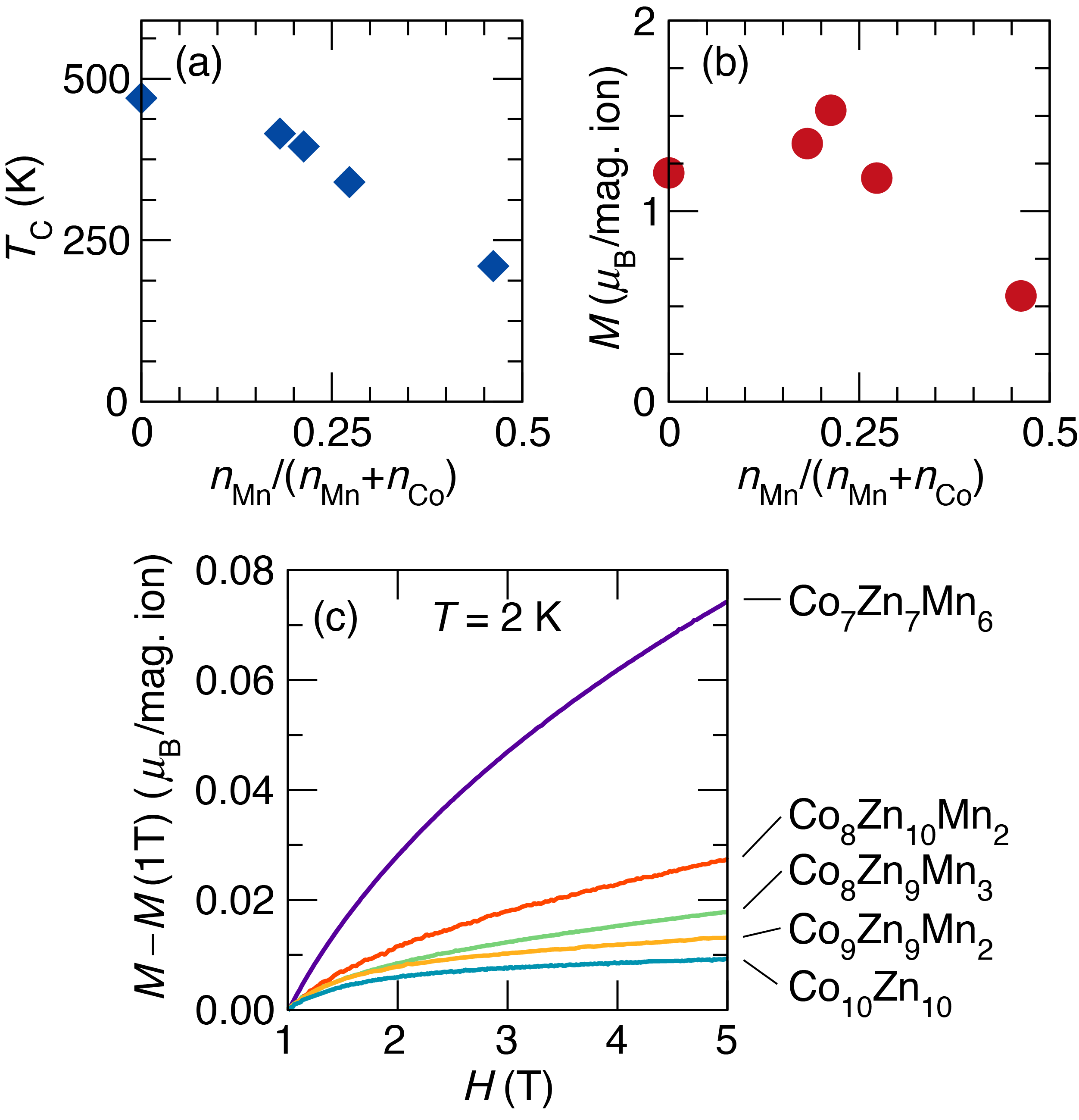}
	\caption{Magnetic properties of Co$_x$Zn$_y$Mn$_z$ alloys derived from the data in Fig.\,\ref{fig:mag-data}. 
	 (a) and (b) show properties as a function of the amount of Mn relative to Co in the sample.
	 (a) With increasing Mn, the Curie temperature drops from 470\,K (\ce{Co10Zn10}) to 210 K (\ce{Co7Zn7Mn6}).
	 (b) Upon addition of Mn into \ce{Co10Zn10}, the average saturated moment on the magnetic ions (Mn and Co) initially increases up until Mn makes up about 20\% of the magnetic ions in the structure. Adding additional Mn results in a decrease in the average moment. The average moments are obtained from the magnetization at $T$\,=\,2\,K and $H$\,=\,5\,T.
	 For (a) and (b), the data points correspond to, in order from left to right, \ce{Co10Zn10}, \ce{Co9Zn9Mn2}, \ce{Co8Zn10Mn2}, \ce{Co8Zn9Mn3}, and \ce{Co7Zn7Mn6}.
	 (c) shows a comparative view of the high-field portions of the 2K. $M(H)$ loops that are shown in Fig.\,\ref{fig:mag-data}. For each sample, the magnetization at 2 K is expressed in units of $\mu_B$ per magnetic ion and the magnetization at $H$\,=\,1\,T is subtracted in order to highlight the size of the moment change experienced by the material at high field. 
	 For \ce{Co10Zn10}, only a small moment change is seen, indicating that the moments in \ce{Co10Zn10} are (very nearly) completely polarized by a 1\,T field. For \ce{Co7Zn7Mn6}, 
	 on the other hand, the moment rises by about 0.08\,$\mu_{\rm{B}}$ per magnetic ion (around 14\%) between 1\,T and 5\,T, indicating that the material still contains unpolarized spins at 1\,T. 
	 Samples with smaller amounts of Mn fall in between \ce{Co10Zn10} and \ce{Co7Zn7Mn6}.
	 }

\label{fig:mag-props}
\end{figure}

While it is known that at low field these materials form various long-wavelength modulated magnetic structures, including helimagnetic, conical, and skyrmion lattice phases,
\cite{Tokunaga2015,Karube2016,Karube2017,Karube2018} the local magnetic structure that is being modulated is not well established. In particular, the magnetic behavior
appears to indicate some glassy character to the magnetism at low temperature \cite{Karube2018}. In this section, we carefully consider magnetization data on the five Co$_x$Zn$_y$Mn$_z$ compositions in light of the atomic distributions determined in the previous sections, building 
up a picture of the local magnetic structure of Co$_x$Zn$_y$Mn$_z$.

Figure~\ref{fig:mag-data} shows DC magnetization data collected on five different Co$_x$Zn$_y$Mn$_z$ compositions. Magnetization as a function of temperature under a low applied field
 ($H$\,=\,20\,mT), displayed in the top row, shows that all five compositions display an onset of magnetic order that looks generally like ferromagnetic ordering, with Curie temperatures ranging from
 470\,K for the most Co-rich sample ({\ce{Co10Zn10}) to 210\,K for the most Mn-rich sample(\ce{Co7Zn7Mn6}).

Below the ordering temperature, however, the susceptibility deviates from that expected for a standard ferromagnet. In particular, in samples with nonzero Mn content, an additional magnetic feature is observed in the low-field susceptibility. Upon cooling to around 120\,K, there is a pronounced downturn in both the ZFC and FC magnetization, as has been previously reported in these compounds \cite{Tokunaga2015,Karube2018}. The magnitude of this downturn seems to generally increase with increasing Mn content, and in the most Mn-rich composition (\ce{Co7Zn7Mn6}), the downturn is accompanied by a large increase in the difference between the ZFC and FC curves. This divergence is concurrent with the onset of magnetic hysteresis as observed in loops of the magnetization as a function of magnetic field (Fig.\,\ref{fig:mag-data}e-ii, 2\,K). Both of these observations point to magnetic irreversibility in \ce{Co7Zn7Mn6} at low temperature. For all other compositions, and for \ce{Co7Zn7Mn6} at higher temperatures, no such hysteresis is observed and a relatively small difference between the ZFC and FC magnetization curves is seen. 
 
Figure~\ref{fig:acms} shows AC magnetic susceptibility for \ce{Co7Zn7Mn6}, the most Mn-rich composition in this study. AC susceptibility as a function of temperature was measured on the sample of 
\ce{Co7Zn7Mn6} with three different excitation frequencies ($f$ = 1\,Hz, 30\,Hz, and 997\,Hz). Above about 70 K, the real part of the susceptibility ($\chi'$) mirrors the field-cooled DC susceptibility shown 
in Fig.\ref{fig:mag-data}e-i. Above the Curie temperature (around 200\,K), the imaginary part of the susceptibility ($\chi''$) is very nearly zero, indicating that the system responds quickly to applied 
magnetic field. This is expected for a magnetic material in its paramagnetic regime. Below 200\,K, a small $\chi''$ signal appears, consistent with the small loss expected due to the onset of magnetic order.

Below 70\,K, the real part of the susceptibility deviates from the field-cooled DC susceptibility, dropping precipitously. Simultaneously, the 
imaginary part of the susceptibility increases, comes to a peak around 30\,K, and then falls off again. As seen on the right part of Fig.~\ref{fig:acms} Both $\chi'$ and $\chi''$ show a clear dependence on 
the excitation frequency---the onset of the downturn in $\chi'$ and the peak in $\chi''$ increase in temperature by about 4.5\,K as the excitation frequency is increased from 1\,Hz to 997\,Hz. These features 
in $\chi'$ and $\chi''$ are a clear hallmark of glassy behavior. 

The shift in freezing temperature as a function of applied frequency gives information about the magnetic interactions involved in the freezing process of a glassy magnetic system \cite{Mydosh2015}. The 
relative shift in freezing temperature per decade of frequency, $\delta T_f = \Delta T_f/ (T_f\Delta \log_{10}(f))$ is 0.038. Canonical spin glasses, such as dilute Mn in Cu or Au, have values of $
\delta T_f$ of the order of 0.005. On the other hand, superparamagnets show much larger values on the order of 0.1 to 0.3. The intermediate value of $\delta T_f$ in \ce{Co7Zn7Mn6} is consistent with 
that of a ``cluster glass", where clusters of spins, rather than individual spins, make up the building blocks of the spin glass. This is the expected state for a spin glass with a large concentration (larger 
than a few percent) of magnetic ions, as is the case with the 12d sublattice of \ce{Co7Zn7Mn6}. 

As can be seen in the magnetization \emph{vs.} field loops, the saturated magnetization of all of the compounds monotonically increases as temperature is lowered, indicating that the low-
temperature downturn seen in the 20 mT $M(T)$ is purely a low-field phenomenon. For \ce{Co8Zn9Mn3} and \ce{Co7Zn7Mn6}, we also show ZFC and FC measurements taken at $H$\,=\,200\,mT (Fig.\,\ref{fig:mag-data}d-e). At this higher field, the low-temperature downturn in susceptibility has nearly disappeared and the irreversibility between ZFC and FC measurements has closed. 

Magnetic properties extracted from the data in Fig.~\ref{fig:mag-data} are shown in table~\ref{tbl:mag} and plotted in Fig.~\ref{fig:mag-props}. As seen in Fig.~\ref{fig:mag-props}, many of the magnetic 
properties follow trends with the relative amount of Mn to Co in the unit cell, quantified as $n_{\rm{Mn}}/(n_{\rm{Mn}}+n_{\rm{Co}}$). The magnetic transition temperature $T_c$ monotonically decreases 
as Mn is added (Fig.\,\ref{fig:mag-props}a), indicating that ferromagnetic exchange is weakening. The average magnetic saturation per magnetic ion (Co and Mn), on the other hand, shows non-
monotonic behavior (Fig.\,\ref{fig:mag-props}b). The moment first rises as small amounts of Mn are 
added to the structure, then falls as larger amounts (3 or more Mn per unit cell) are added (Fig.\,\ref{fig:mag-props}). This behavior, at least at first glance, appears to disagree with the density 
functional theory calculations, which indicate that each Co should contribute a moment of about 1.3 $\mu_B$ while each Mn contributes a moment of about 3.2 $\mu_B$ (or 2.4 $\mu_B$ if it is on 
its minority 8c site). Based on these calculations, we would expect that addition of Mn will uniformly increase the saturated magnetic moment. For \ce{Co10Zn10}, \ce{Co9Zn9Mn2} and 
\ce{Co8Zn10Mn2}, the measured magnetic moment per formula unit is reasonably consistent with these local moment values (within 15\%). For the more Mn-rich compositions 
\ce{Co8Zn9Mn3} and \ce{Co7Zn7Mn6}, the experimental saturated moment is far smaller than predicted.

This apparent disagreement can be explained by non-collinearity of the Mn spins. The DFT calculations considered only collinear ferromagnetic moments; therefore, while the calculations highlight the importance of a large local moment on the Mn atoms (energy scale on the order of eV), they do not necessarily inform the lowest energy orientation of those moments (energy scale on the order of meV). Here, we propose an understanding of the magnetic structure of Co$_x$Zn$_y$Mn$_z$ that is consistent with both the DFT and the magnetic measurements. In our model, \ce{Co10Zn10} behaves as a local ferromagnet, with Co spins collinearly aligned within a unit cell. When a small number of Mn atoms are added to this system, they hold, as predicted by the DFT, a large local moment. These local moments order via the dominant exchange field established by the surrounding Co moments and tend to align ferromagnetically with the Co moments, increasing the susceptibility and saturation magnetization of the system. As more Mn is added, Mn atoms become more likely to have Mn-rich neighborhoods and therefore are less subject to the ferromagnetic exchange field established by the Co moments. The Mn moments tend less toward order, and more towards the native behavior of Mn in the $\beta$-Mn structure, which has a disordered ground state. Therefore, we see that after the addition of about two Mn atoms per unit cell, the saturated moment begins to decline with the addition of Mn.

In this picture, the Mn spins are more-or-less dynamically disordered at temperatures below the apparent ordering temperature of the Co moments, with a partial tendency to align with the Co spins. The low-temperature downturn in susceptibility, then, can be understood as glassy freezing of clusters of the Mn spins into a statically disordered spin glass, at which point the Mn spins no longer respond to a small magnetic field. In the most Mn-rich sample, this transition is even accompanied by a dramatic increase in magnetic hysteresis, consistent with the onset of sluggish magnetic dynamics in the frozen state.

Figure~\ref{fig:mag-props}c shows a detailed view of the high-field portion of the magnetization as a function of field for each sample at 2\,K with the magnetization at 1\,T subtracted off to highlight differences in the high-field magnetic behavior. In this high-field regime, the chiral magnetic modulations are believed to be completely field-polarized such that no long-wavelength structures persist. \ce{Co10Zn10} experiences very little change in moment between 1\,T and 5\,T. The total moment in samples containing Mn, on the other hand, increases with application of large magnetic fields. In particular, the moment of \ce{Co7Zn7Mn6} rises about 0.08 $\mu_B$/mag. ion (14\%). This suggests that the moments are not fully polarized at 1\,T, or even 5\,T, consistent with our model of disordered Mn on the 12d site.

\subsection{Sublattice-specific magnetic ordering}

Based on the magnetization data, we have inferred that the Mn and Co moments behave very differently in Co$_x$Zn$_y$Mn$_z$ materials. In order to clarify this behavior, low-temperature neutron diffraction measurements were performed
on the \ce{Co8Zn9Mn3} and \ce{Co7Zn7Mn6} samples. The long-wavelength helimagnetic groundstate will, in principle, give a neutron diffraction pattern that looks very much like
the diffraction pattern of the unmodulated unit cell, except with magnetic satellite peaks around some of the magnetic Bragg peaks. In this case, however, the modulation period is so long that the expected
reciprocal-space splitting will be too small to see in the resolution of our powder neutron diffraction experiment. Because of this, we are able to model the neutron
diffraction patterns as if the long-period modulation does not exist, refining for only the local magnetic structure.

Figure~\ref{fig:893-mag-riet} shows low-temperature neutron diffraction patterns, along with Rietveld fits, for the \ce{Co8Zn9Mn3} sample. The site occupancies have been fixed from the paramagnetic (350\,K) synchrotron-neutron corefinement. The 100\,K pattern is corefined with a synchrotron pattern taken at the same temperature (shown in Supplemental 
Fig.\,S1a), while the 14\,K pattern is refined on its own. Upon cooling through the magnetic transition, the intensity of several peaks is increased, and two new peaks at low $Q$ (1.4 \AA$^{-1}$ and 3.67 \AA$^{-1}$) 
appear. These two peaks can be indexed as the allowed peaks (110) and (111) in the $\beta$-Mn unit cell, which have a very low structure factor in the paramagnetic phase. No magnetic peaks are observed 
at structurally-forbidden positions, indicating that the magnetic unit cell and the paramagnetic unit cell are the same.

For magnetic Rietveld (co)refinements on this sample, a simple collinear ferromagnetic spin structure was assumed. In this model, the Co and Mn on the 8c site have a single, linked moment, and the Co and Mn on the 
12d site have a different linked moment. This means that the refinement is probing the average ordered moment of the Co and Mn on the 8c and 12d sites. As can be seen in Fig.\,\ref{fig:893-mag-riet}, this model fits the 
neutron patterns at both 100\,K and 14\,K data well, with no discripencies in peak intensities between the model and the magnetic peaks in the data. At 100\,K, the 8c magnetic ions are found to hold moments of 1.00(6) 
$\mu_B$ and the 12d magnetic ions are found to hold moments of 0.99(28) $\mu_B$ (table~\ref{tbl:mag-refinements}). For the 8c site, which is mainly comprised of Co, this value is of reasonable agreement with the 
DFT moment (about 1.3\,$\mu_B$). On the 12d site, however, the magnetic ions are mostly Mn, and this refined moment is smaller than the DFT moment (3.3\,$\mu_B$). This is consistent with our understanding of the 
Mn spins as dynamically disordered, with only a partial tendency to align in the direction of the Co spins at 100\,K.

At 14\,K, below the glass freezing temperature, similar magnetic diffraction is observed. A close look at the pattern reveals that the magnetic intensity is actually weaker in the 14\,K pattern than the 100\,K pattern. Such behavior is unusual in ferromagnets, where at low temperature fluctuations are suppressed and the observed ordered moment is at its largest. As seen in Table~\ref{tbl:mag-refinements}, refining the 14\,K pattern gives moments on the 8c and 12d site of 0.80(8)\,$\mu_B$ and $0.1(4)$\,$\mu_B$, respectively. The moments on both sites have decreased relative to the 100\,K data, but the moment on the Mn-rich 12d site has decreased much more than that on the Co-rich 8c site. This supports our hypothesis that the frustrated and glassy behavior observed in the magnetization data is driven by the Mn spins on the 12d site, while the Co atoms on the 8c site remain largely ferromagnetic. At 100K, the Mn spins may be dynamically disordered with some preference to align themselves with the ferromagnetic exchange field of the Co atoms, while at 14\,K the disordered Mn spins are frozen into a glassy state with no long-range alignment or net moment. The minor drop in moment on the 8c site may be attributed to the small amount of Mn-Co antisite mixing, which was noted earlier.

\begin{figure}[thb!]
	\centering
	\includegraphics[width=\columnwidth]{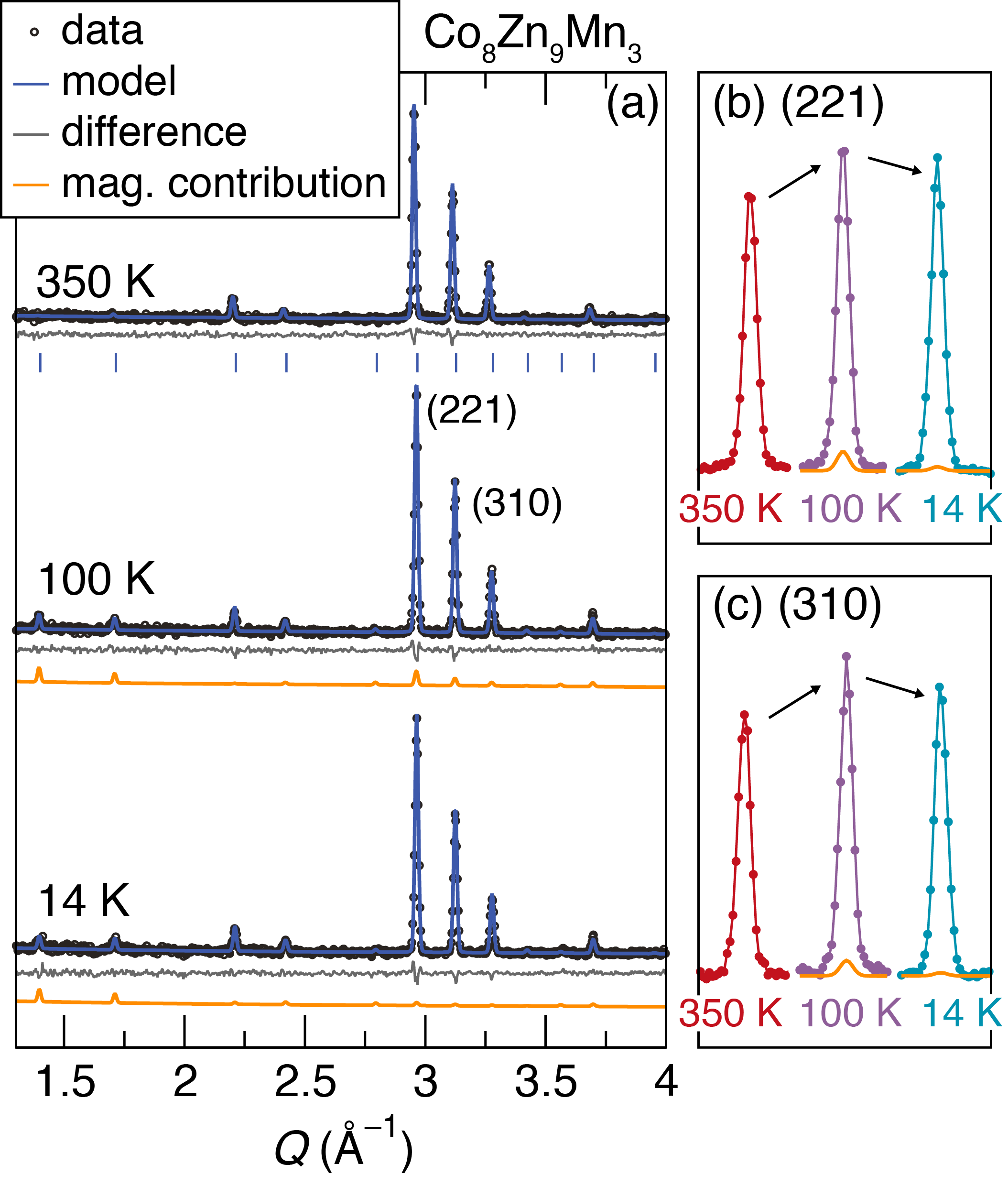}
	\caption{Magnetic neutron Rietveld refinements for the \ce{Co8Zn9Mn3} sample at 100\,K and 14\,K. (a) shows how a magnetic contribution to the pattern grows in below the magnetic ordering temperature. Tick marks underneath the 350\,K pattern indicate allowed nuclear peaks for the $\beta$-Mn structure; no magnetic intensity is seen at forbidden peak positions, however some magnetic peaks occur at allowed structural peak positions which have nearly zero intensity in the paramagnetic pattern. (b) and (c) shows a close view of the temperature evolution of the largest structural peaks, the (221) and (310). In both cases, the peak is seen to have a larger intensity at 100\,K than 14\,K, driven by the larger magnetic contribution (orange) at 100\,K than 14\,K. }
	\label{fig:893-mag-riet}
\end{figure}

\begin{figure}[thb!]
	\centering
	\includegraphics[width=\columnwidth]{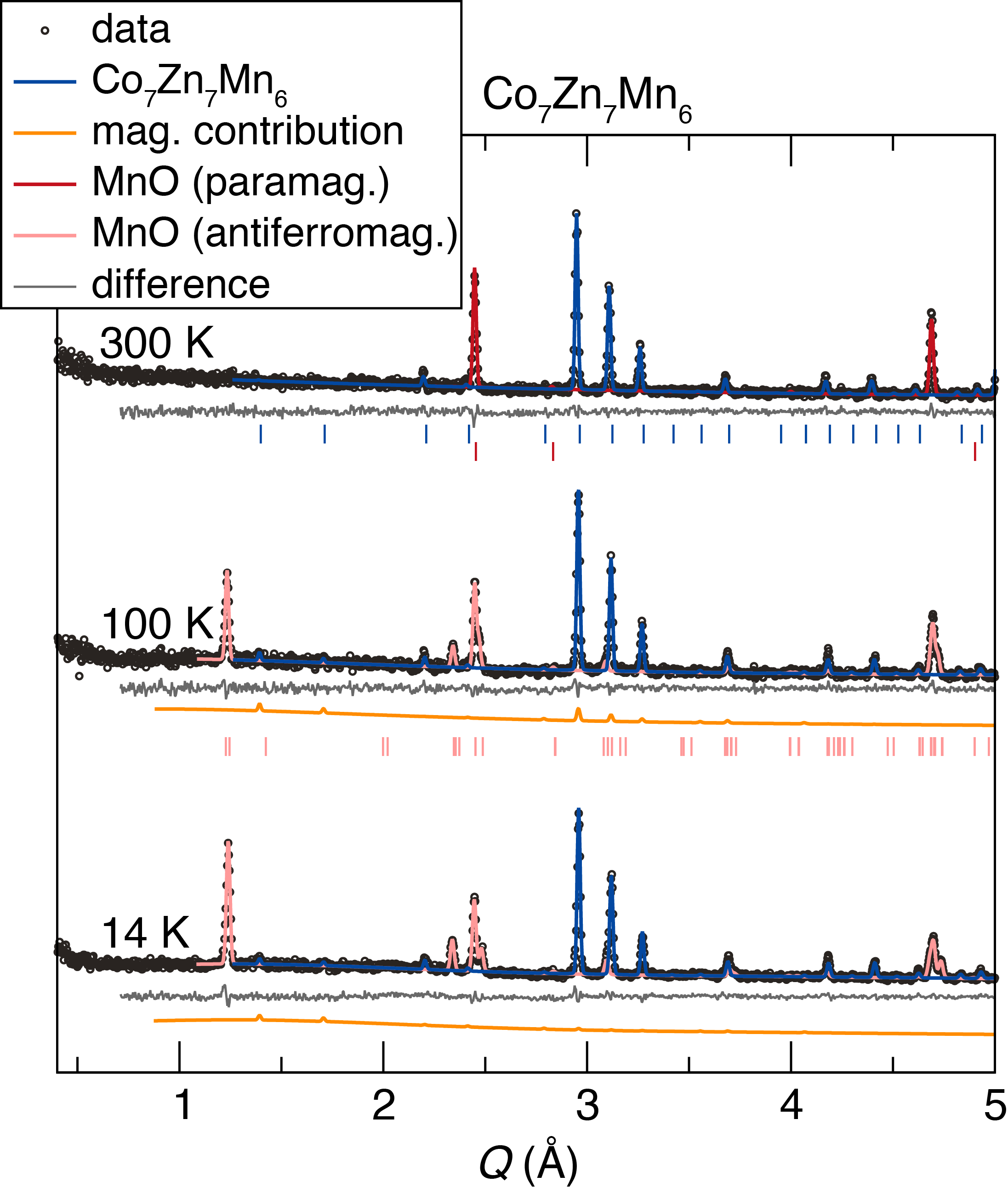}
	\caption{Magnetic neutron Rietveld refinements for the \ce{Co7Zn7Mn6} sample at 100\,K and 14\,K. Below the observed magnetic ordering temperature (210\,K), a small magnetic contribution is found to grow in to the $\beta$-Mn structure. The paramagnetic to antiferromagnetic transition of MnO at 118\,K is also seen. This transition is accompanied by a rhombohedral structural distortion, which, along with the magnetic ordering, changes the space group from $Fm\overline{3}m$ to the monoclinic group $C_c2/c^\prime$ \cite{Frandsen2015}.}
	\label{fig:776-mag-riet}
\end{figure}

\begin{table*}[!htb]
\centering
\caption{Results of magnetic Rietveld refinement of neutron diffraction data. $M_{\rm{sat}}$ is determined DC magnetization data. Numbers in parentheses are standard uncertainties in the last given digit(s) from Rietveld refinement.}
\begin{tabular}{@{\extracolsep{5pt}}lccccccc}
\toprule
               & $T$ &  \multicolumn{2}{c}{mom. per mag ion ($\mu_B$)} & total moment & $M_{\rm{sat}}$ ($H$\,=\,2\,T) & $\chi^2$ ($r_{\rm{wp}}$/$r_{\rm{exp}}$) \\ \cline{3-4} 
               & (K)     &   8c     & 12d      & ($\mu_B$/f.u.) & ($\mu_B$/f.u.) &                  \\ \colrule
\ce{Co8Zn9Mn3} & 350     &         &          &          &             & 0.84 (6.12/6.66) \\                                          
               & 100   & 1.00(6) & 0.99(28) & 11(1)    & 12.1        & 0.98 (6.39/6.47) \\                                          
               & 14     & 0.82(8) & 0.15(41) & 7.0(1.4) & 12.3 & 0.98 (6.52/6.60) \\                                          
\ce{Co7Zn7Mn6} & 300     &         &          &          &             & 0.80 (4.28/4.79) \\                                          
               & 100    & 0.64(5) & 0.75(16) & 6.3(1.0) & 6.7         & 0.85 (4.44/4.81) \\   
               & 14      & 0.38(4) & 0         &  3.0(3)            & 7.3   &  1.05 (3.59/3.51) \\
\botrule  
\end{tabular}
\label{tbl:mag-refinements}
\end{table*}

In the \ce{Co7Zn7Mn6} sample (Fig.\,\ref{fig:776-mag-riet}), the magnetic contribution to the 100\,K pattern is even smaller than in the \ce{Co8Zn9Mn3}. A Rietveld refinement assuming collinear ferromagnetism gives a 
moment of 0.64(5) $\mu_B$ per magnetic ion on the 8c site and 0.75(16) $\mu_B$ per magnetic ion on the 12d site (Table~\ref{tbl:mag-refinements}). These small ordered moments suggest that, as seen in the 
magnetization data, the \ce{Co7Zn7Mn6} sample shows more magnetic disorder than the \ce{Co8Zn9Mn3} sample. Nevertheless, at this temperature, both the Co spins and the Mn spins apparently have some tendency 
to order, as seen by the nonzero refined moment, which is consistent with the saturated magnetic moment at 100\,K (6.3(1.0) $\mu_B$ \emph{vs.} 6.7$\mu_B$). Like in \ce{Co8Zn9Mn3}, the magnetic contribution to the 
pattern drops as the sample is cooled through the observed glass transition. At 14\,K, the moment on the 12d site refined to zero, and so was fixed to zero while the 8c site was allowed to refine to 0.38(4) $\mu_B$. This 
suggests that, as was the case in \ce{Co8Zn9Mn3}, the 12d site is completely magnetically disordered at low temperature while the Co-rich 8c site maintains some net order. 

In a conventional spin glass, the spins transition directly from a fluctuating paramagnetic state to a frozen glass state as temperature is lowered. Here, we see that the Mn spins in Co$_x$Zn$_y$Mn$_z$ go through at least three regimes. At the observed Curie temperature of the material, the moments transition from a paramagnetic state at high temperature to a dynamically fluctuating, yet partially-ordered state, at intermediate temperatures.  At lower temperature, the spins freeze into a fully disordered spin glass, with no net moment. Based on this behavior, we may characterize these materials as``reentrant'' spin cluster glass, as has been observed in some other systems \cite{Bao1999,Miyazaki1988,Dho2002,Chatterjee2009}. An interesting feature of such systems is that, because the system transitions from a spin glass state to a ``ferromagnetic'' state, upon warming, the ``ferromagnetic'' state must have higher entropy than the glass state, despite its being an ordered state. In this case, this apparent contradiction is resolved by the fact that the ferromagnetic state is in fact a dynamic, fluctuating state which is only partially ordered, and therefore has a higher entropy than the frozen spin glass.

\subsection{Nature of the magnetic transitions in Co$_x$Zn$_y$Mn$_z$}

\begin{figure}[t!]
	\centering
	\includegraphics[width=\columnwidth]{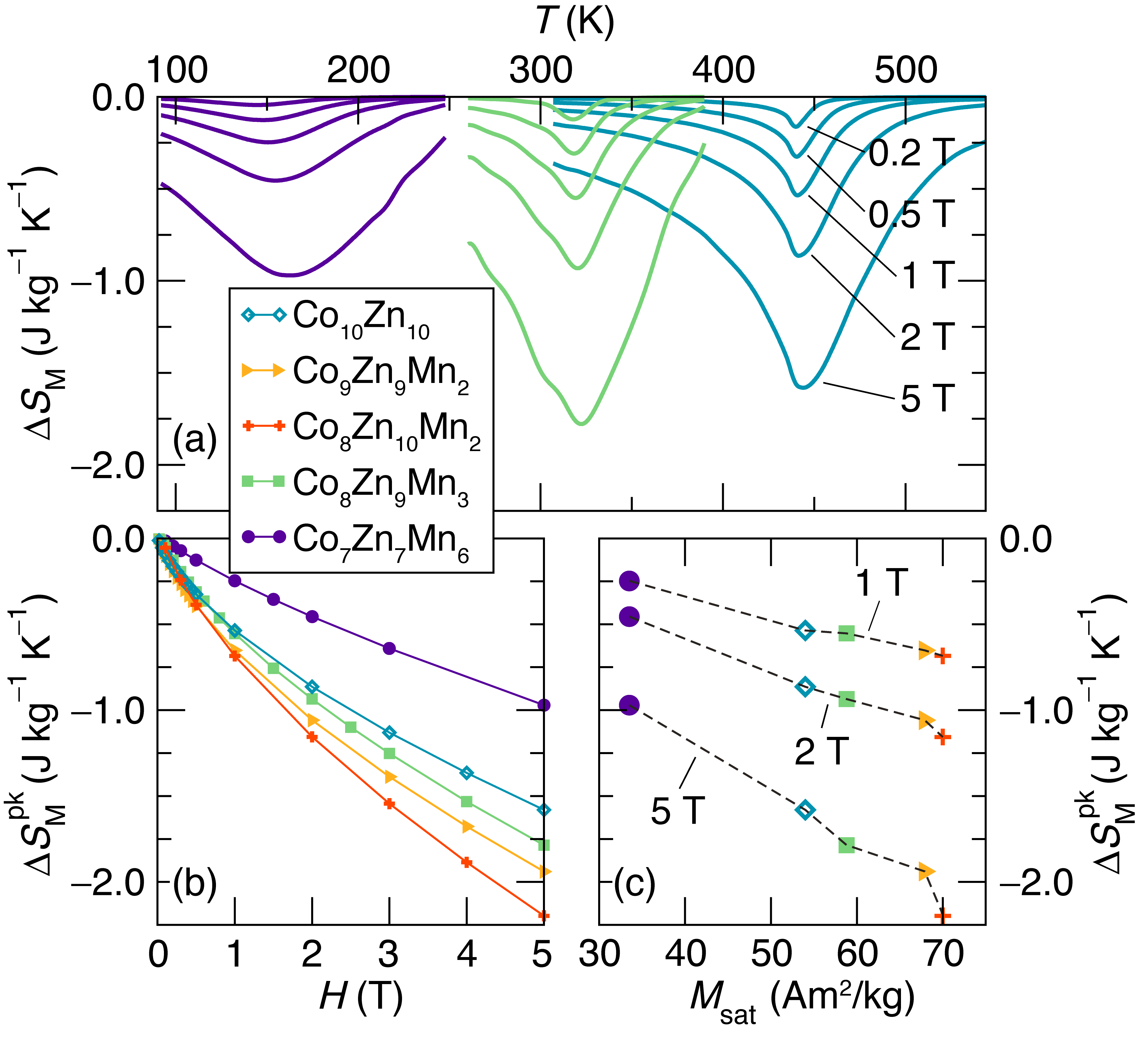}
	\caption{Magnetocaloric evaluation of Co$_x$Zn$_y$Mn$_z$. $\Delta S_M (T, H)$, the entropy change of the material upon isothermal magnetization to magnetic field $H$, is calculated from magnetization data using equation 
	\ref{eqn:deltaSm}. Larger $|\Delta S_M|$ values indicate a larger magnetocaloric effect. (a) $\Delta S_M$ as a function of temperature for several applied field values for three Co$_x$Zn$_y$Mn$_z$ compositions. The largest effect is 
	seen at the Curie temperature for each material. (b) Peak values of $\Delta S_M$ as a function of applied field for five Co$_x$Zn$_y$Mn$_z$ compositions. (c) Peak $\Delta S_M$ for different Co$_x$Zn$_y$Mn$_z$ compositions as a function of magnetic saturation of the composition. There is a clear relationship between magnetocaloric effect and magnetic moment in these samples.}
	\label{fig:DSm}
\end{figure}

\begin{figure}[htb]
	\centering
	\includegraphics[width=\columnwidth]{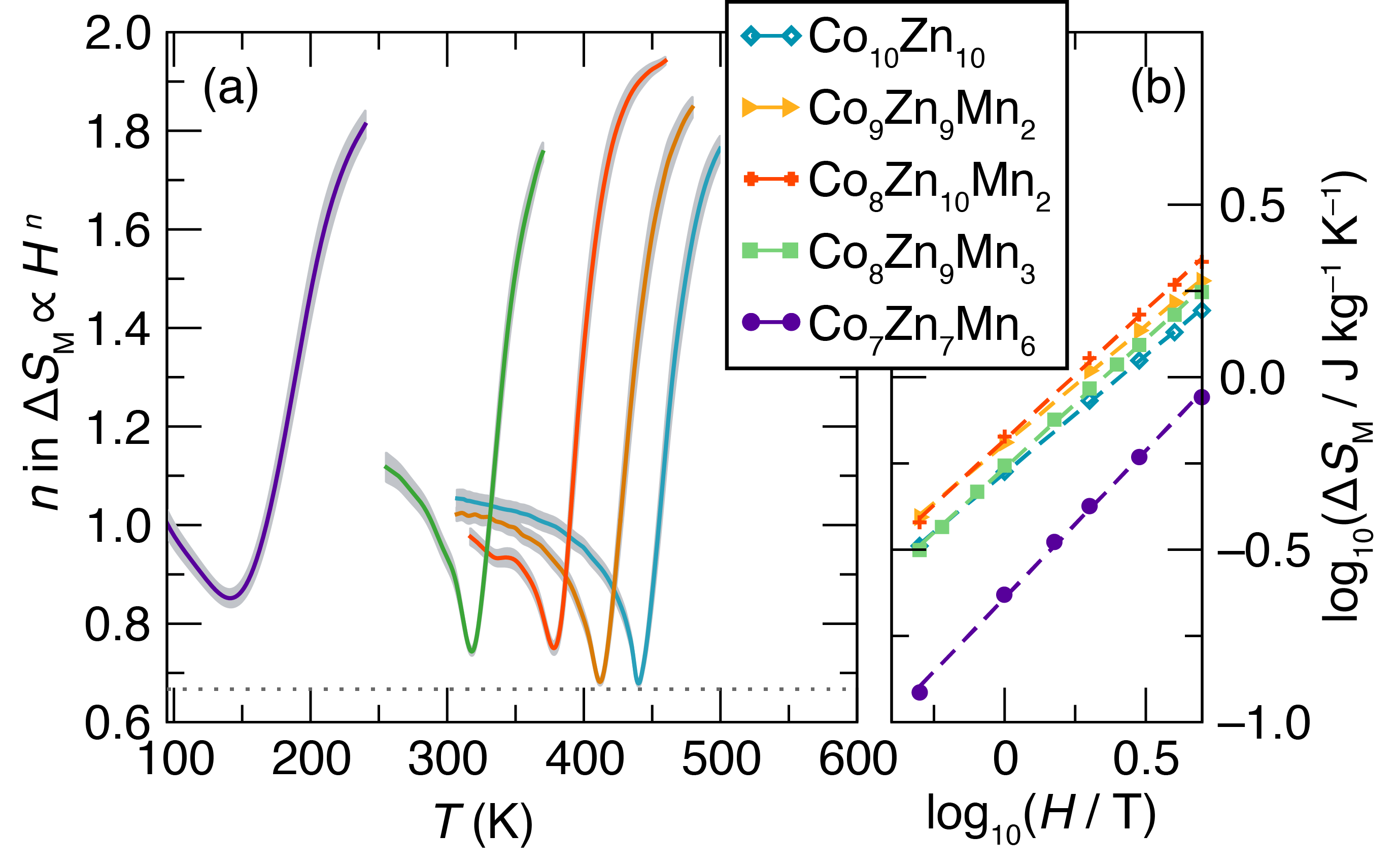}
	\caption{(a) Temperature dependence of the power law exponent, $n$, of the magnetocaloric effect ($\Delta S_M(H) \propto H^{n}$). This exponent is obtained from linear fits of log-log plots of $\Delta S_M$ \emph{vs.} field for applied fields between 500 mT and 5T, as shown for the critical temperatures in (b). For all samples, the power law exponents are found to show a standard shape, characteristic of a continuous magnetic transition. The minimum $n$ in each curve is the critical exponent $n_c$, which occurs at the critical temperature $T_c$. The dashed line shows the expected critical exponent for the mean field model, $n_c = 2/3$. As Mn content increases, $n_c$ increases relative to the mean field model value, suggesting increasingly disordered magnetic interactions.}
	\label{fig:power-law}
\end{figure}

In order to further understand the effect of the Mn moment disorder on the magnetic properties of Co$_x$Zn$_y$Mn$_z$ materials, we characterized the magnetocaloric properties
of the samples using indirect isothermal entropy change measurements, which can be obtained from DC magnetization data. Isothermal magnetic entropy upon application of fields up to 5\,T is shown 
in Fig.~\ref{fig:DSm} for the five samples studied. In each case, the dominant effect is negative, \emph{i.e.} application of a magnetic field decreases the entropy of the system, as is typical for a 
ferromagnet, where an applied field suppresses spin fluctuations. As expected, the largest negative effect is observed near the observed Curie temperature, where the spins are most easily polarized by an external field.

As can be seen in Fig.~\ref{fig:DSm}c, the peak magnetocaloric effect is observed to be directly related to the low-temperature saturation magnetization, which in turn is related non-monotonically to the composition (Fig.~\ref{fig:mag-data}b). While saturation magnetization is not, in general, a good predictor of magneticaloric effect \cite{Bocarsly2017}, it is often the case that $\Delta 
S_M^{pk}$ correlates with saturation magnetization within a specific family of materials \cite{Franco2012, Tishin1997, Mandal2004, Singh2016, Levin2017},
particularly in cases where there are no structural transitions concurrent with the magnetic transition. 
 
Using this $\Delta S_M$ data, we have performed an analysis of power law exponents associated with this magnetic entropy change, $\Delta S_M(H) \propto H^{n}$. At the critical 
point temperature $T_c$ this exponent is a critical exponent ($\Delta S_M(H, T=T_c) \propto H^{n_c}$). For a continuous 
magnetic transition, this $n_c$ can be related to the other magnetic critical exponents of the system by applying equation \ref{eqn:deltaSm} to the Arrott-Noakes equation of state \cite{Arrott1967, Franco2006}, 
yielding:
\begin{equation} 
\label{eqn:crit}
n_c = 1 + \frac1\delta\left(1-\frac1\beta\right)
\end{equation}
$\delta$ and $\beta$ are the standard critical exponents relating $H$ to $M$ and $M$ to $T$, respectively, and are readily available for various of magnetic exchange models, giving $n_c = 2/3$ for the 
mean-field model, and lower values for Ising, Heisenberg, and XY models. Beyond the critical behavior, it has been found that the power law relation $\Delta S_M \propto H^{n}$ often holds across a broad 
temperature range, with $n$ showing a distinctive temperature evolution, increasing from $n_c$ to 1 as temperature is decreased and increasing from $n_c$ to 2 at the temperature is increased. 

As seen in Fig.\,\ref{fig:power-law}, this canonical shape of $n_c$ is observed for all Co$_x$Zn$_y$Mn$_z$ samples when considering only the high field data ($H \ge $0.5\,T). The minimum exponent ($n_c$), which is 
seen at the critical temperature $T_c$, however, evolves as Mn is added. For CoZn, the exponent is 0.68(1), which is comparable to the mean field theory value. As Mn is added, $n_c$ rises 
monotonically reaching 0.85(2) for \ce{Co7Zn7Mn6}. Any values of $n_c$ larger than 2/3 cannot be explained with the standard models of ferromagnetism (mean field, Heisenberg, XY, and Ising). 
However, elevated values are generally obtained for ferromagnetic bulk magnetic glasses \cite{Franco2006,Franco2006a,Franco2008a,Franco2011,Kaul1985}, where magnetic 
interactions are disordered and multi-scale. Taken together, Figs.\,\ref{fig:DSm} and \ref{fig:power-law} show that the Co$_x$Zn$_y$Mn$_z$ samples behave, at fields above about 0.5\,T, as ferromagnets with continuous transitions and 
reduced effective moments due to the fluctuating Mn spins. These Mn spins serve to progressively disorder the overall magnetic interactions, including those felt by the Co atoms.

\begin{figure*}[!htb]
	\centering
	\includegraphics[width=0.9 \textwidth]{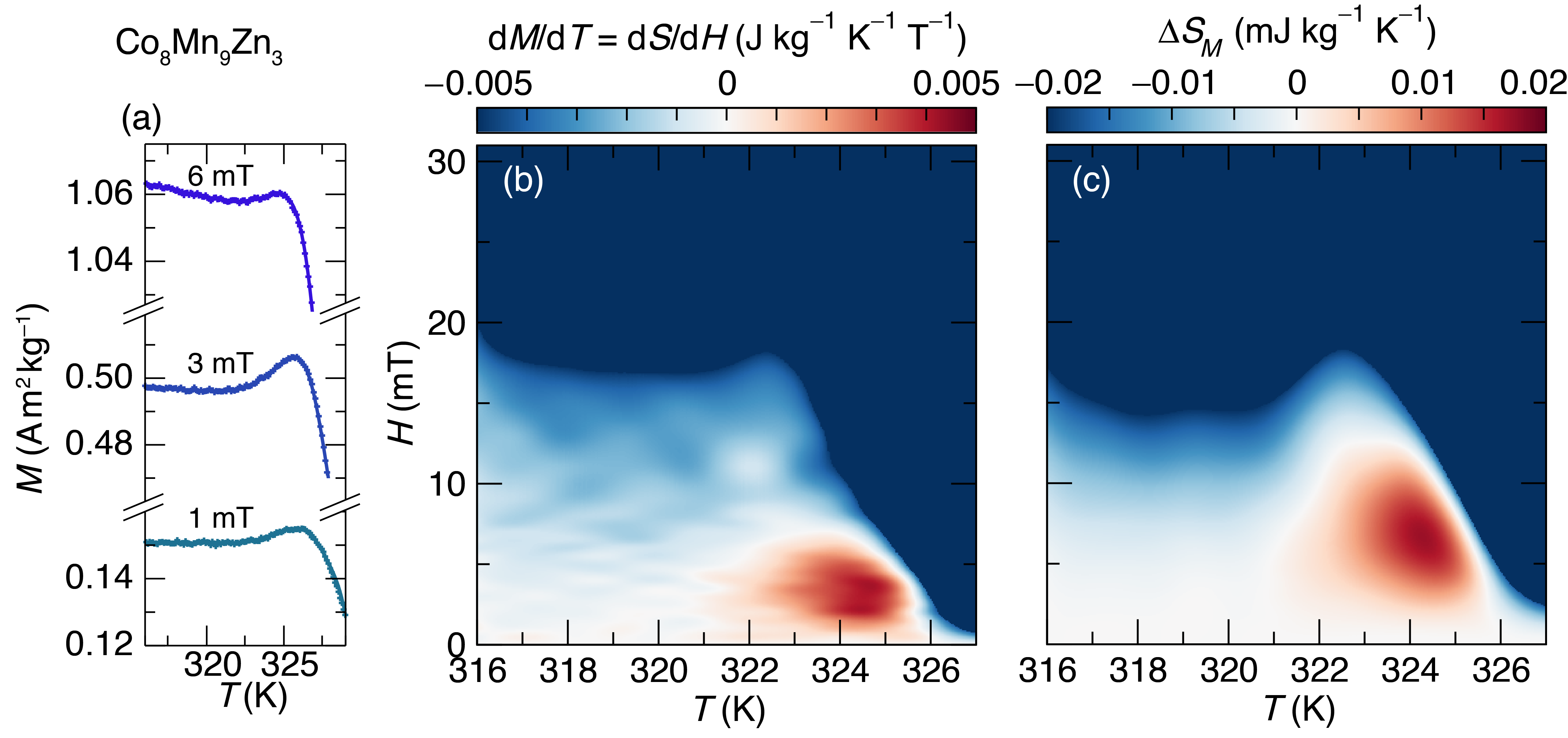}
	\caption{Magnetoentropic analysis of the low-field behavior of \ce{Co8Zn9Mn3}. (a) Representative $M$ \emph{vs.} $T$ data under three different applied fields. The slope of many such curves, $\partial M/\partial T = \partial S/\partial H$ is shown as a heatmap in (b), with colorbar chosen symmetrically about zero to highlight the areas of conventional (negative, blue) and inverse (positive, red) magnetocaloric effect. (c) Heatmap of the isothermal entropy change upon application of a given magnetic field at a given temperature $\Delta S_M(T,H)$, which is obtained by integrating (b) in the field direction (eqn.~\ref{eqn:deltaSm}).  This analysis reveals three clear areas: (i) A blue region at high temperature and field representing the region in which the sample is behaving as a conventional field-polarized ferromagnet or paramagnet. In this regime, application of a magnetic field suppresses fluctuations and decreases entropy. (ii) A white region at low temperatures and fields, where the system shows a long-period helimagnetic or partially polarized conical magnetic phase. In this region, application of a magnetic field causes the planes of spins in the conical structure to progressively cant in the direction of the field, but has little effect on the entropy of the system. (iii) A red pocket just below the Curie temperature. This corresponds to the phase region where a hexagonal skyrmion lattice has been observed in this class of materials and other cubic skyrmion hosts. The increased entropy in this region is due to increased entropy in the skyrmion lattice relative to the helical or conical phases. However, due to the disorder in the sample, it is difficult to resolve this feature into clear phase boundaries, and differentiate it from a Brazovskii transition.}
\label{fig:magnetoentropy}
\end{figure*}

The lower-field magnetocaloric effect is expected to be more complex in skyrmion hosts materials as the various first order phase transitions between the skyrmion lattice phase and other chiral and non-chiral phases 
manifest themselves as anomalies in the magnetization and magnetocaloric effect \cite{Bocarsly2018}. Figure~\ref{fig:magnetoentropy} shows a detailed measurement of the low-field magnetocaloric effect 
in a 0.74\,mg piece of \ce{Co8Zn9Mn3}. A region of positive (red) $\Delta S_M$, indicating high entropy relative to the conical, helical, or ferromagnetic phases, can be seen in the field-
temperature phase diagram just below the magnetic ordering temperature, where the equilibrium skyrmion lattice is expected. This is consistent with our understanding of the skyrmion lattice as a dynamic, 
fluctuation-stabilized phase. In the case of FeGe, this anomalous magnetocaloric behavior can be resolved into a phase diagram involving first-order transitions between the skyrmion lattice phase and 
the surrounding conical phase and a separate first-order Brazovskii transition \cite{Brazovskii1975,Janoschek2013} between the ordered phases and a fluctuation-disordered phase at higher 
temperatures \cite{Bocarsly2018}. In the case of our Co$_x$Zn$_y$Mn$_z$ samples, these individual phase transition lines are difficult to resolve due to macroscopic disorder in the samples. Supplemental Material Fig.
\,S4 shows a comparison of the $M(T)$ of a 66\,mg piece and a 0.74\,mg piece of the same \ce{Co8Zn9Mn3} sample. The smaller piece shows a clear bump (precursor 
anomaly) in the magnetization just beneath the onset of magnetic ordering, while the larger sample shows no such anomaly. This bump, when processed through eqn.~\ref{eqn:deltaSm}, becomes the 
anomalous magnetocaloric signal associated with the skyrmion lattice formation. Apparently, in larger samples, the subtle anomaly is completely smeared out by compositional variation across the sample, or by an inhomogenous demagnetizing field. Even in the 0.74\,mg sample, some smearing of this feature occurs relative to the signal seen in small single crystals of well-ordered metal-metalloid compound like FeGe or MnSi.

\section{Conclusions}
The Co$_x$Zn$_y$Mn$_z$ system of high-temperature skyrmion hosts presents an interesting case of coexistence of several types of magnetic interactions. Like previously studied skyrmion host materials, Co$_x$Zn$_y$Mn$_z$ shows a low-field phase diagram made up of a variety of long-wavelength ($\lambda$ between 100\,nm and 200\,nm) modulated magnetic structures, including several types of skyrmion lattices. 
Here, we have carefully characterized the atomic and magnetic disorder in Co$_x$Zn$_y$Mn$_z$ materials with various compositions, explaining the exotic local magnetic structure that those chiral modulations are built upon.

Through synchrotron and neutron diffraction, as well as DFT total energy calculations, we find that both the 8c and 12d site of Co$_x$Zn$_y$Mn$_z$ have atomic disorder. Neutron and synchrotron diffraction, as well as DFT calculations, 
show that the 8c atomic sublattice is mostly filled with Co atoms, while the larger 12d site is mostly filled with Mn and Zn. Mn's stability on the large 12d site is driven by its ability to develop a large local 
moment on that site. 

DC and AC Magnetic measurements, and magnetic neutron diffraction refinements clarify the disordered magnetic behavior of these materials. Co$_x$Zn$_y$Mn$_z$ shows two-sublattice magnetic behavior, 
with coexistence of small, ferromagnetic Co moments behaving like a conventional ferromagnet, and large, dynamically disordered Mn moments which continue fluctuating below the ordering temperature of the 
Co spins, and ultimately freeze into a reentrant clustered spin glass at low temperature while the Co spins remain largely ordered. This behavior is observed in all of our samples containing Mn, with the magnetic 
signatures of disorder and glassiness increasing with increasing Mn concentration. 

This unique magnetic structure allows for the dramatic and novel skyrmionic phase behavior recently observed in the Co$_x$Zn$_y$Mn$_z$ system. For example, in \ce{Co8Zn8Mn4}, a metastable skyrmion lattice is 
observed down to low temperatures, including below the magnetic glass transition temperature \cite{Karube2017}. In \ce{Co7Zn7Mn6}, a new type of equilibrium disordered skyrmion lattice is found near the glass 
transition temperature \cite{Karube2018}. In both cases, the observations are attributed to the influence of disorder. The two-sublattice magnetic structure allows for the coexistence of ordered  magnetism (on the Co atoms) and
disordered (on the Mn atoms) magnetism.  This means that long-range ordered skyrmion lattice phases may coexist with, and be influenced by, a disordered magnetic glass system.

\section{Acknowledgements}
   This work was supported by  the National Science Foundation through the MRSEC Program of the National Science Foundation through DMR-1720256 (IRG-1). J.D.B. is supported by the NSF Graduate Research Fellowship Program under Grant No. 1650114. We thank Dr. Weiwei Xie for helpful insights and Dr. Neil Dilley for help with the high-temperature magnetic measurements.  Use of the Advanced Photon Source at Argonne National Laboratory was supported by the U. S. Department of Energy, Office of Science, Office of Basic Energy Sciences, under Contract No. DE-AC02-06CH11357. We thank the 11-BM staff for their assistance with the data collection. We acknowledge the support of the National Institute of Standards and Technology, U. S. Department of Commerce, in providing the neutron research facilities used in this work. We also acknowledge the use of the facilities of the Center for Scientific Computing at UC Santa Barbara.
  
Certain commercial equipment, instruments, or materials are identified in this paper to foster understanding. Such identification does not imply recommendation or endorsement by the National Institute of Standards and Technology, nor does it imply that the materials or equipment identified are necessarily the best available for the purpose.

\end{document}


\title{Supplemental Material: Deciphering structural and magnetic disorder in the chiral skyrmion host materials
Co$_x$Zn$_y$Mn$_z$ ($x+y+z=20$)}

\author{Joshua D. Bocarsly}
\email{jdbocarsly@mrl.ucsb.edu}
\affiliation{Materials Department and Materials Research Laboratory, 
University of California, Santa Barbara, California 93106, United States}

\author{Colin Heikes}
\affiliation{Center for Neutron Research, National Institute of Standards and Technology,\\
Gaithersburg, Maryland 20899, United States} 

\author{Craig M. Brown}
\affiliation{Center for Neutron Research, National Institute of Standards and Technology,\\ 
Gaithersburg, Maryland 20899, United States} 

\author{Stephen D. Wilson}
\affiliation{Materials Department and Materials Research Laboratory, 
University of California, Santa Barbara, California 93106, United States} 

\author{Ram Seshadri}
\affiliation{Materials Department and Materials Research Laboratory, 
University of California, Santa Barbara, California 93106, United States}

\maketitle

\section{Additional diffraction patterns}
\begin{figure}[!h]
	\centering
	\includegraphics[width=0.5\columnwidth]{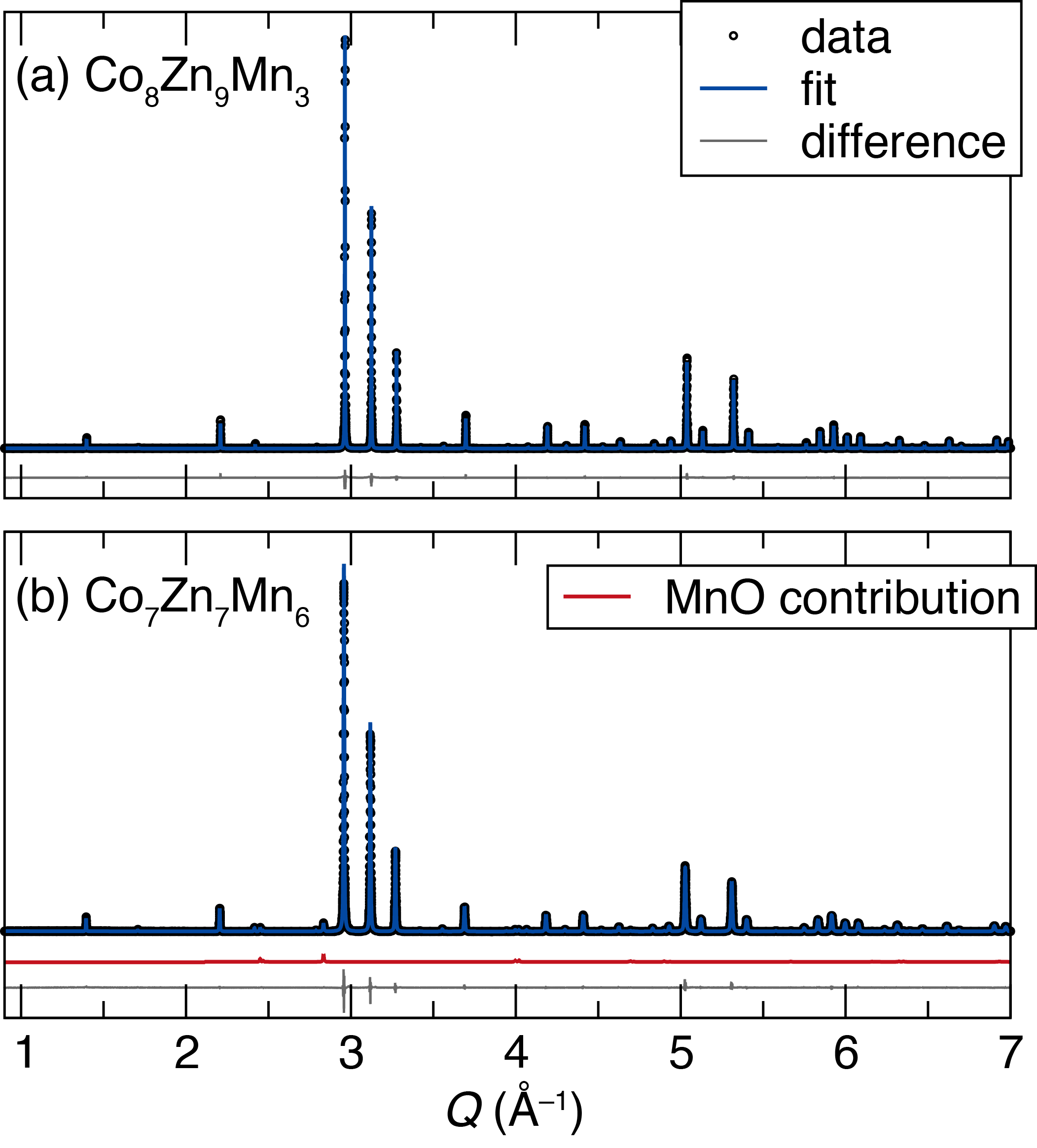}
	\caption{Synchrotron powder diffraction of \ce{Co8Zn9Mn3} and \ce{Co7Zn7Mn6} at 100K, displayed with fit from Rietveld co-refinement with neutron data shown in the main text, Fig.\,7.
	}
	\label{fig:sync-100K}
\end{figure}

\begin{figure}[!h]
	\centering
	\includegraphics[width=0.5\columnwidth]{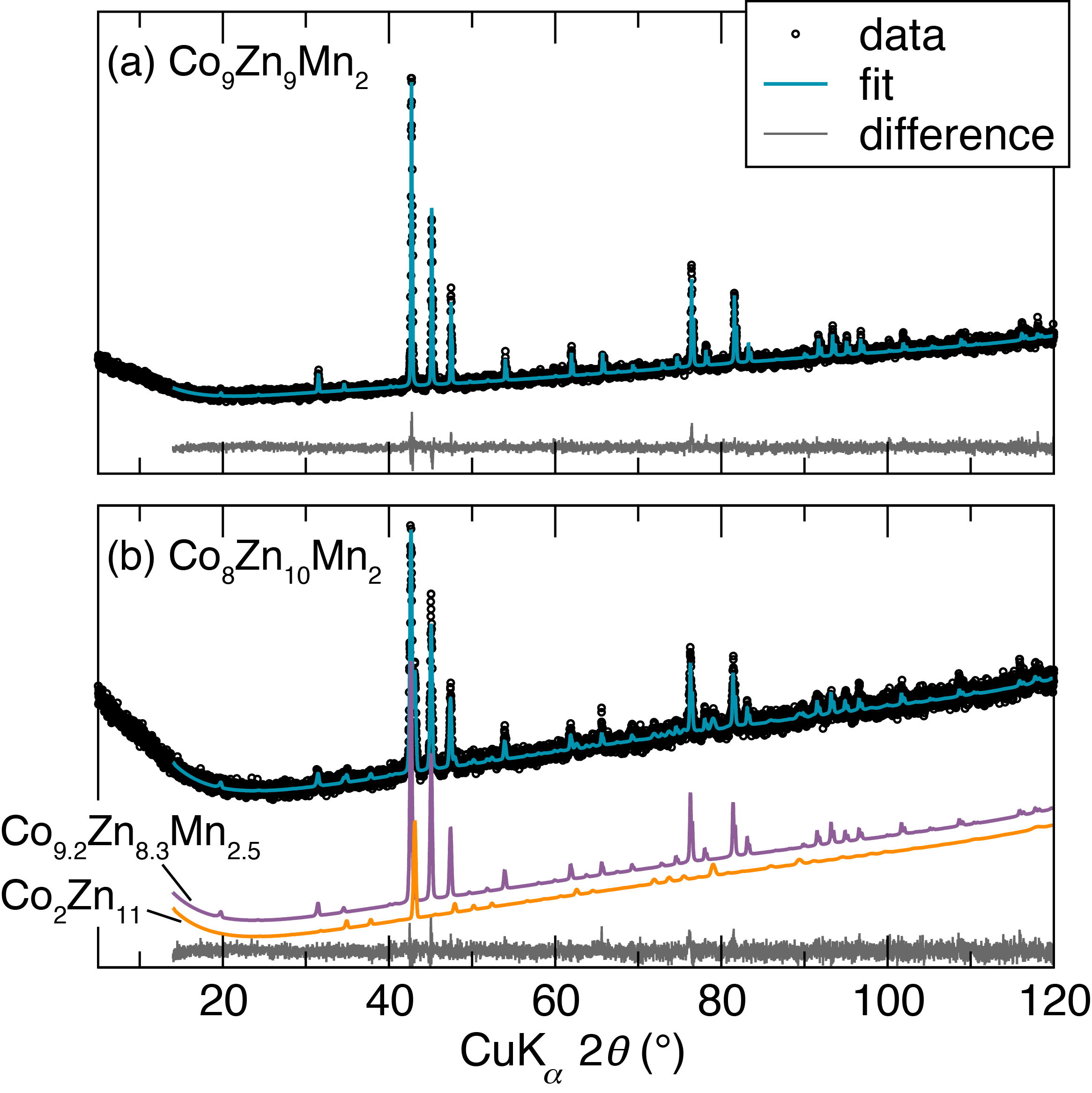}
	\caption{Laboratory X-ray diffraction of \ce{Co9Zn9Mn2} (a) and \ce{Co8Zn10Mn2} (b), displayed with Rietveld fit. The \ce{Co9Zn9Mn2} sample shows no resolvable impurities, while the \ce{Co8Zn10Mn2} samples contains 20.1(6)\,wt.\% of the $\beta$-brass Co$_{9.2}$Zn$_{8.3}$Mn$_{2.5}$
	}
	\label{fig:lab-xrd}
\end{figure}

\section{Magnetization \emph{vs.} temperature for C\lowercase{o}$_x$Z\lowercase{n}$_y$M\lowercase{n}$_z$ samples}

\begin{figure}[!h]
	\centering
	\includegraphics[width=0.75\columnwidth]{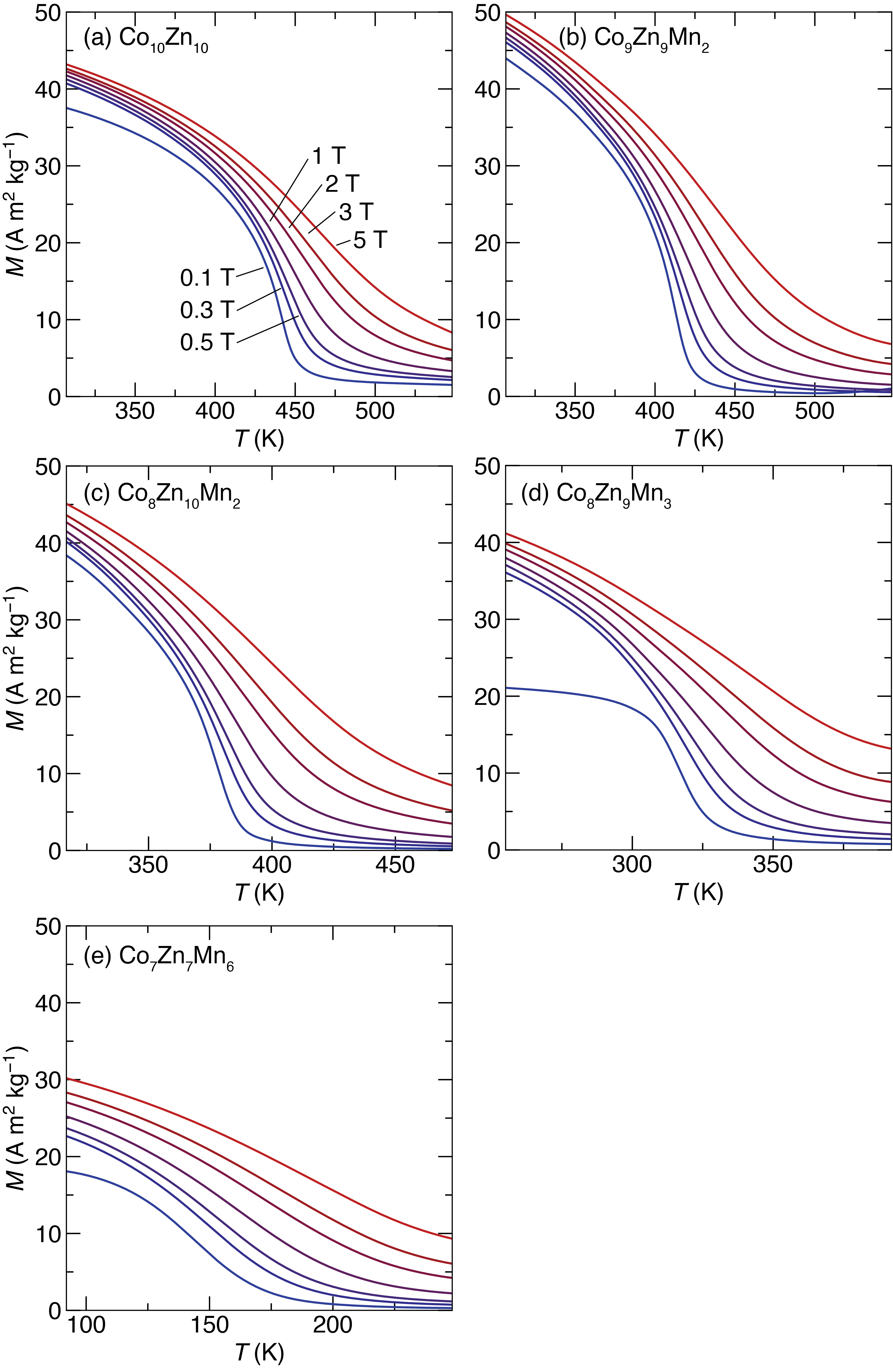}
	\caption{Magnetization as a function of temperature at different applied fields for each of the samples. This data is processed into the $\Delta S_M$, and its power law exponent, as displayed in the main text Figs.\,8-9.}
	\label{fig:all_MTs.png}
\end{figure}

\section{Sample size dependence of precursor anomaly in C\lowercase{o}$_8$Z\lowercase{n}$_9$M\lowercase{n}$_3$}

\begin{figure}[!h]
	\centering
	\includegraphics[width=0.6\columnwidth]{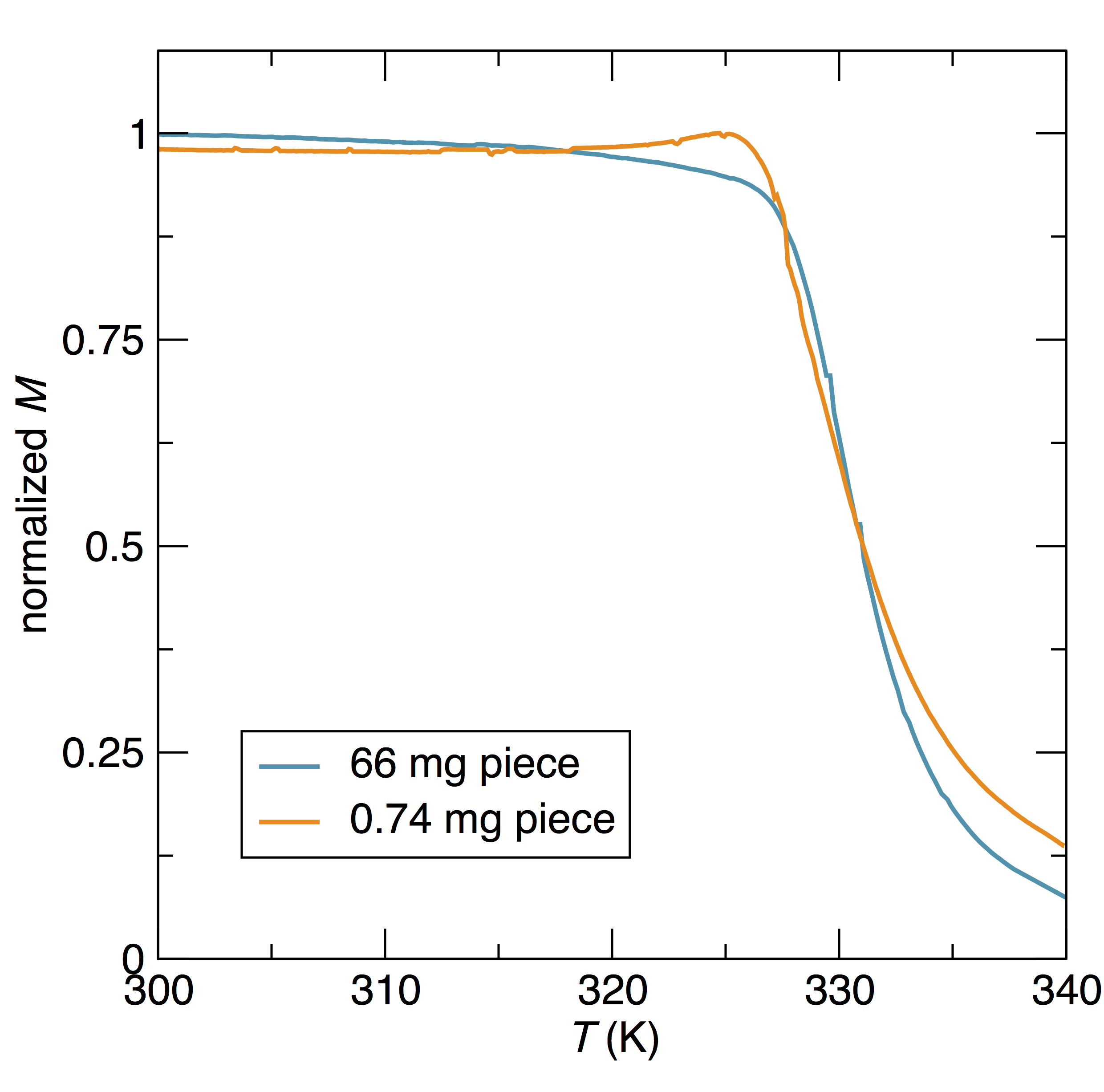}
	\caption{Comparison of magnetization \emph{vs.} temperature under an applied field of 5\,mT for two different pieces taken from a single sample of \ce{Co8Zn9Mn3}. The smaller piece shows a clear precursor anomaly- a small bump below the Curie temperature- while the larger piece does not. This may be caused by compositional disorder across the sample, or an inhomogeneous demagnetizing field. The larger sample contains a range of different magnetic transition temperatures, smearing out the subtle magnetic precursor anomalies. The smaller sample shows comparatively clearer behavior.}
	\label{fig:anomaly}
\end{figure}